\documentclass[%
 reprint,
superscriptaddress,
showpacs,%preprintnumbers,
 amsmath,amssymb,
 aps,
]{revtex4-2}
\usepackage[T1]{fontenc}
\usepackage[utf8]{inputenc}
\usepackage{rotating}
\usepackage{stackengine}
\usepackage{graphicx}
\usepackage{color}
\usepackage{verbatim}
\usepackage{subfigure}
\usepackage{tikz}
\usepackage{tkz-graph}

\definecolor{Blue}{rgb}{0,0,0}
\definecolor{Red}{rgb}{1,0,0}
\definecolor{Black}{rgb}{0,0,0}
\definecolor{Green}{rgb}{0.0, 0.5, 0.0}

\newcommand{\DBR}[1]{{\color{Blue} #1}}
\newcommand{\SM}[1]{{\color{Blue} #1}}

\begin{document}

\preprint{APS/123-QED}

\title{\DBR{Robust} continuous-variable multipartite entanglement in circular \DBR{arrays} of nonlinear waveguides} 

\author{Sugar Singh Meena} 
\email{sugar15$_$sps@jnu.ac.in}
\affiliation{School of Physical Sciences, Jawaharlal Nehru University, New Delhi 110067, India}
\author{David Barral} 
\email{david.barral.rana@gmail.com}
\affiliation{Galicia Supercomputing Center (CESGA), Avda.\ de Vigo S/N, Santiago de Compostela, 15705, Spain}
\affiliation{Quantum Materials and Photonics Research Group, Optics Area, \\Department of Applied Physics, iMATUS,
Faculty of Physics, Faculty of Optics and Optometry, \\University of Santiago de Compostela, Santiago de Compostela, 15872, Spain.}
\author{Ankan Das Roy} 
\affiliation{School of Physical Sciences, Jawaharlal Nehru University, New Delhi 110067, India}
\author{Sunita Meena} 
\affiliation{School of Physical Sciences, Jawaharlal Nehru University, New Delhi 110067, India}
\author{Amit Rai} 
\email{amitrai@mail.jnu.ac.in }
\affiliation{School of Physical Sciences, Jawaharlal Nehru University, New Delhi 110067, India}

\begin{abstract}
Encoding continuous-variable quantum information in the optical domain has recently enabled the generation of large entangled states, yet robust implementation remains a challenge. Here, we present a straightforward protocol for generating multipartite entanglement based on spontaneous parametric down-conversion in a circular array of quadratic nonlinear waveguides. We provide a rigorous theoretical framework, including comprehensive derivations of the propagation equations and the identification of regimes where analytical solutions are possible. \DBR{Crucially, our approach identifies the pump and detection configurations required to sustain and measure multipartite full inseparability across arbitrary propagation distances and for any number of waveguides $N=4 n$. This regime, elusive to standard numerical methods, represents a key requirement for scalable quantum protocols. Our} scheme is inherently robust as it relies on phase-matched propagation eigenmodes, making it resilient against variations in sample length, coupling, and nonlinearity.
\end{abstract}

\maketitle

\section{Introduction}\label{I}

Multipartite entanglement is a cornerstone of quantum information technologies, underpinning advances in communication, sensing, and computing \cite{cacciapuoti2024multipartite, zhuang2018distributed, santra2025genuine}. As the scale of these entangled resources grows, so does the complexity of the tasks that can be performed. Large-scale cluster states—the essential ingredient for measurement-based quantum computation (MBQC)—have already been realized in bulk optics using path encoding and various forms of multiplexing \cite{yukawa2008experimental, cai2017multimode,  armstrong2012programmable, larsen2019deterministic, asavanant2019generation, jia2025continuous, roh2025generation}. However, while integrated photonic chips offer a promising route toward scalable and stable hardware, generating spatial multimode entanglement on these platforms remains a significant challenge.

Optical quantum information is typically encoded in discrete variables (DV), such as path or polarization degrees of freedom. While DV encodings provide near-perfect gate fidelity, they are fundamentally hindered by probabilistic photon sources and cryogenic detection \cite{li2022chip, chen2021quantum}. In contrast, continuous variable (CV) encoding via quadrature amplitudes allows for deterministic quantum state generation—a capability that has been leveraged for secure communication, error correction, and quantum-enhanced sensing \cite{hajomer2024long, larsen2021fault, ge2025heisenberg}; and room temperature detection \cite{kouadou2025homodyne, tasker2021silicon}. Particularly, 
nonlinear waveguide circuits can confine, interfere with, and detect light on a chip, and provide stability and scalability essential for complex CV optical quantum systems \cite{masada2015continuous, lenzini2018integrated}. This goal seemed far from feasible with bulk optics-inspired schemes, but scaling enabled by these waveguide circuits could lead to the development of usable optical devices \SM{\cite{moody20222022, pelucchi2022potential, clark2025integrated}}.

\DBR{
The planar arrays of linear waveguides offer a promising solution to the stability and scalability challenges \cite{yang2025programmable}. However, these systems often remain limited by their reliance on external bulk optical setups, typically spontaneous parametric downconversion (SPDC) sources, for state generation. In contrast, arrays of nonlinear waveguides (ANWs) solve this issue by integrating quantum state generation and manipulation within a single device \cite{christodoulides2003discretizing, iwanow2004observation, setzpfandt2009competing, solntsev2012spontaneous}. In the DV regime, successful experimental demonstrations of ANWs \cite{solntsev2014generation, raymond2024tunable, raymond2025tailoring} have sparked extensive theoretical research for a deeper insight and applications in quantum technologies \cite{ belsley2020generating, hamilton2022quantum, he2024optimizing, delgado2025transport}. Although previous works have predominantly addressed planar waveguide arrays, other topologies introduce novel functionalities. In particular, circular arrays are pivotal for applications such as optical switching, power distribution, and quantum state transfer \cite{longhi2007light, hudgings2002design, hizanidis2006localized, rai2022transfer}. Moreover, multidimensional entanglement has been demonstrated using multicore fibers driven by external optical sources \cite{lee2018generation, gomez2021multidimensional}. 

In the CV regime, multipartite entanglement in planar waveguide arrays has been analyzed in different regimes and through different methods: from second-harmonic generation \cite{barral2019zero} to SPDC \cite{barral2020versatile}, and from numerical methods \cite{rai2012dynamics} to analytical solutions \cite{barral2021scalable}. Importantly, analytical solutions for ANWs leverage certain symmetries between array eigenmodes, or supermodes, and pump profiles. When these symmetries are exploited, multipartite entanglement appears in planar ANWs with an odd number of waveguides via its zero supermode, i.e. the eigenmode with a zero eigenvalue that is phase-matched throughout propagation yielding entanglement between its odd-indexed individual components \cite{barral2019zero, barral2021scalable}. Recently, multimode entanglement has also been explored in squeezed-vacuum-injected circular arrays of linear waveguides \cite{anuradha2024production}, and in circular ANWs for up to six modes \cite{meena2025theoretical}. However, this approach for studying circular ANWs relies primarily on numerical simulations for small number of waveguides, and insight on key aspects such as entanglement scalability and the entanglement structure of the states created in these nonlinear devices are still missing.

In this work, we introduce a practical and straightforward protocol for generating and characterizing an important set of multipartite CV entangled states of light in circular arrays of $\chi^{(2)}$ nonlinear waveguides. Our approach produces analytical solutions for a circular ANWs with any number of waveguides $N=4n \,(n=1,2,\dots)$ in the SPDC regime. Our contributions are three-fold. First, while planar ANWs possess a single zero supermode, circular ANWs exhibit a unique feature: they host two  supermodes, or Fourier modes, of zero eigenvalue. Under suitable pumping, these modes are efficiently squeezed along propagation, generating two decoupled sets of entangled states composed of even and odd individual components, respectively. Consequently, circular arrays double the number of entangled modes compared to planar ANWs. Perhaps surprisingly, despite the strong direct coupling between physically adjacent waveguides, the resulting quantum states form two separate, interlaced sets of spatially entangled modes. Second, based on three unique discrete symmetries of Fourier modes, we derive orthonormality relations that can be exploited in ANWs through suitable pumping. Hence, three different pump phase profiles produce unique output correlations that can be calculated analytically. Crucially, this setup allows for an active switching of entanglement, offering a functional \emph{entanglement switch} tailored for quantum networking applications. Third, our analytical solutions yield the set of pump profiles and measurement bases that produce a robust entanglement independently of the number of waveguides --being a multiple of four-- and the propagation length. In contrast, numerical methods fail to determine these pump profiles and detection bases, yielding an oscillatory behaviour --periodically transitioning between entangled and separable states-- dependent on the number of waveguides and the propagation length \cite{meena2025theoretical}. The analytical solutions are thus helpful to design and fabricate experimental devices for entanglement generation without relying on specific values of the array parameters: number of modes, coupling, nonlinearity, and length.  

Thus, by exploiting discrete topological symmetries of circular arrays, pump phase profiles, efficient build-up of zero Fourier modes, and analytical solutions, the nonlinear circular array turns into an efficient, versatile, and robust tool for the generation of multipartite CV entanglement. 
}

We have organized the article as follows: We introduce the circular ANWs and study the dynamics of SPDC light in the array in Sec. \ref{II}. We then discuss the Fourier modes of a linear circular array and their orthonormality relations in Sec. \ref{III}. Using these relations, we present analytical solutions for the circular ANWs with uniform phase pump profile, alternating $\pi$-phase and $\pi/2$-phase pump profile in Sec. \ref{IV}. Using the solution for the uniform phase profile, we analyze the generation of multipartite entanglement \DBR{and its robustness to losses} in Sec. \ref{V}. In Sec. \ref{VII}, we discuss the feasibility and future extensions of our work. Finally, we conclude in Sec. \ref{VIII}.

\begin{figure}[t]
  \centering
    \includegraphics[width=0.45\textwidth]{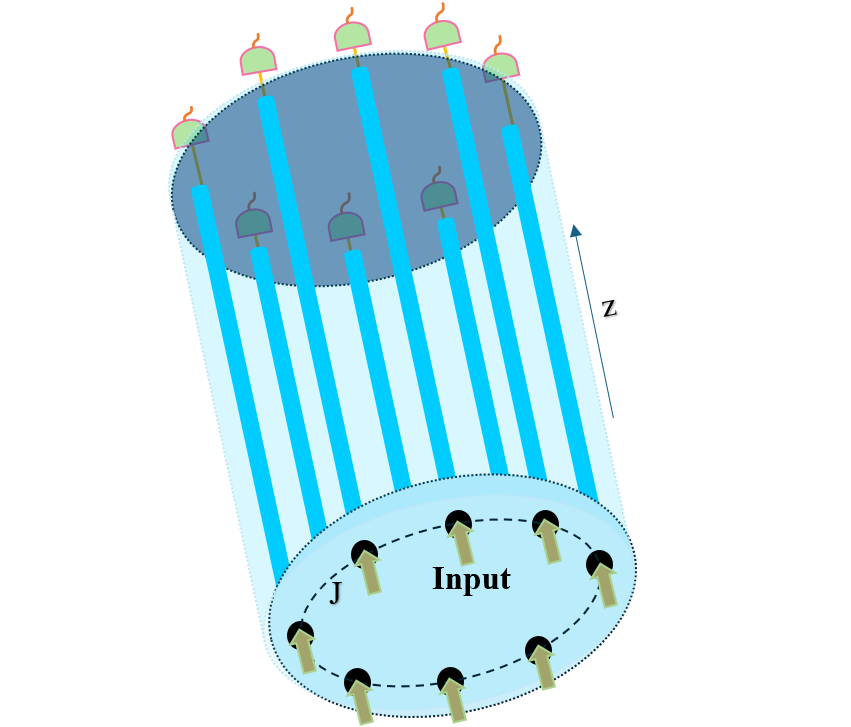}
\caption{\label{F1}\small{Sketch of a nonlinear circular array made up of eight waveguides working in a SPDC configuration, pumping all the waveguides. Nearest-neighbor waveguides are evanescently coupled through the coupling constant $J$. Quantum noise variances and correlations can be measured by multimode balanced homodyne detection. $z$ is the propagation direction.}}
\end{figure}

\section{Dynamics of the circular array of nonlinear waveguides}\label{II}
The circular array consists of \textit{N} identical single-mode second-order $\chi^{(2)}$ nonlinear waveguides in which degenerate SPDC takes place, and the generated fields interact through the nearest-neighbor evanescent couplings as illustrated in Figure \ref{F1}. The coupling between non-neighboring waveguides can often be neglected due to the negligibly weak coupling coefficients \cite{somekh1973channel}. In each waveguide of the circular array, an input harmonic field at frequency $\omega_{h}$ undergoes type-0 down-conversion (identical polarization modes) to generate a signal field at frequency $\omega_{s}$ \cite{alibart2016quantum}. The type-I and type-II down-conversion can also be analyzed similarly; however, we focus on type-0 here. We assume that the phase-matching condition $\Delta \beta \equiv \beta(\omega_{h}) - \SM{2 \beta(\omega_{s})} = 0$, where $ \beta(\omega_{h,s})$ is the propagation constant at frequencies $\omega_{h}$ and $\omega_{s}$, is satisfied only in the coupling zone. Furthermore, we consider that the pump is a strong coherent classical field, which is sufficiently strong to prevent depletion, such that its amplitude can be taken as a parameter. The photons of the signal modes propagating in each waveguide are exchanged between the coupled waveguides through evanescent coupling. In contrast, the evanescent coupling between the pump modes is negligible due to their high confinement in the waveguides for the considered propagation length \cite{noda1981design}. The underlying physical process taking place in $\chi^{(2)}$ nonlinear waveguides can be described by a Momentum operator. \DBR{While the temporal evolution of an optical field is ruled by the Hamiltonian obtained by integration over the differential volume $dx dy dz$ of the quantized $T^{00}$ element (energy density) of the energy-momentum stress tensor $T$ of the electromagnetic field, the propagation of an optical field along the spatial axis $z$ is generated by a Momentum operator obtained by integration over the hyperplane $c dt dx dy$ of the quantized $T^{33}$ element (flux of momentum density) of $T$ \cite{horoshko2022generator}. In this case, we obtain}
\begin{equation}
\label{one}
\hat{\mathcal{M}} =  \hbar \sum_{j=1}^{N} \eta_{j} [ (\hat{\mathcal{A}}_{j}^\dagger)^2 + \hat{\mathcal{A}}_{j}^2 ] + \hbar J_{j} \sum_{j=1}^{N} [\hat{\mathcal{A}}_{j}^\dagger \hat{\mathcal{A}}_{j+1}+\hat{\mathcal{A}}_{j} \hat{\mathcal{A}}_{j+1}^{\dagger}].
\end{equation}  
The following Heisenberg equation is obtained for a circular array of \textit{N} evanescently coupled $\chi^{(2)}$ nonlinear waveguides in the SPDC regime 
\begin{equation}
\label{two}
    \frac{d\hat{\mathcal{A}}_j}{dz} = -2 i \eta_{j} \hat{\mathcal{A}}_{j}^{\dagger} - i J_{j} ( \hat{\mathcal{A}}_{j-1} +  \hat{\mathcal{A}}_{j+1}),
\end{equation}
where $N+j \equiv j~ (mod ~N)$, i.e. $\hat{\mathcal{A}}_{N} \equiv \hat{\mathcal{A}}_{0}$, $\hat{\mathcal{A}}_{N+1} \equiv \hat{\mathcal{A}}_{1}$, and $j = 1,2,..., N$ is the individual mode index. The individual mode basis operators $\hat{\mathcal{A}}_{j} \equiv \hat{\mathcal{A}}_{j} (z, \omega_{s})$ represent the monochromatic slowly varying amplitude annihilation operators of signal (\textit{s}) photons in the $j^{th}$ waveguide, where $[\hat{\mathcal{A}}_{j} (z, \omega),  \hat{\mathcal{A}}_{j^{'}}^{\dagger} (z, \omega^{'})] = \delta{(\omega - \omega^{'})} \delta_{j,j^{'}}$. The effective nonlinear coupling constant of the $j^{th}$ waveguide is given by $\eta_{j} = g\alpha_{h,j}$, where nonlinear constant \textit{g} is proportional to the $\chi^{(2)}$ coefficient and to the spatial overlap of the signal and harmonic fields within each waveguide, and $\alpha_{h,j}$ is the undepleted strong coherent pump field propagating in the $j^{th}$ waveguide. The parameter $\eta_{j}$ can be fine-tuned by adjusting a suitable set of pump phases and amplitudes at each waveguide. $J_{j}$ is the linear coupling constant between modes \textit{j} and \textit{j+1}, and z is the coordinate along the propagation direction. Both coupling and non-linear constants are functions of the signal frequency $J\equiv J(\omega_{s})$ and $g\equiv g(\omega_{s})$; we set them as real without loss of generality. We focus on the relevant case of homogeneous coupling along the propagation, i.e., \textit{J} does not depend on \textit{z}.

The array of nonlinear waveguides offers a large number of adjustable parameters that can be engineered to fit specific applications and operational needs. These customizable parameters in a compact device are a key strength of ANWs compared to other approaches for generating multimode entangled states. In this paper we will focus on a single tunable parameter which can be dynamically adjusted: the pump phase profile $\phi_{j}=arg(\alpha_{h,j})$. As we show below, this paper demonstrates how to utilize this tunable parameter to generate different \DBR{regimes of multipartite entanglement and even to switch it on and off}.

We are interested in continuous-variable features of this system, so we will use along the paper the field quadratures $x_{j}$ and $y_{j}$, where $x_{j} = \hat{\mathcal{A}}_{j} + \hat{\mathcal{A}}_{j}^{\dagger}$ and $y_{j} = i(\hat{\mathcal{A}}_{j}^{\dagger} - \hat{\mathcal{A}}_{j})$ are, respectively,  the orthogonal amplitude and phase quadratures corresponding to a signal optical mode $\hat{\mathcal{A}}_{j}$. The value of shot noise is 1 in this notation. The system of Equations (\ref{two}) can be rewritten in a compact form in terms of the individual-mode quadratures.
\begin{equation}
\label{three}
    \frac{d\hat{\xi}}{dz} = \Delta \hat{\xi},
\end{equation}
where $\hat{\xi} = (\hat{x}_{1},\hat{y}_{1},..., \hat{x}_{N},\hat{y}_{N})^{T}$ and $\Delta$ is a $2N \times 2N$ matrix of coefficients.

We first need to solve the propagation equation to understand the dynamics of light in the ANWs. Equations (\ref{two}) or (\ref{three}) can be solved numerically for a given set of parameters $J_j$, $\eta_{j}$, and \textit{N}, or even analytically for small \textit{N}. However, as the number of waveguides \textit{N} increases, the system becomes more complex, making it difficult to gain physical insight from numerical or low-dimensional analytical solutions. \SM{The problem of propagation in circular ANWs is simplified by using the eigenmodes of the corresponding circular array of \textit{N} waveguides in the linear regime} with $\eta_{j} \propto g = 0$ — the linear or propagation supermodes $\mathcal{B}$. Notably, for circular arrays, these supermodes are discrete Fourier modes \cite{hudgings2002design}. As we will see, the properties of Fourier modes are particularly useful to find analytical solutions. In the next section, we introduce some important properties of discrete Fourier modes.

\section{\label{III} Fourier modes of a linear circular array}

A circular array of \textit{N} linear waveguides with coupling only between nearest-neighbor waveguides is represented by an \DBR{$N\times N$ symmetric circulant matrix $\mathrm{J}$ with elements $\mathrm{J}_{j,~ j+1 ~ (mod ~N)} = J_{j}$. From now on, we consider a homogeneous coupling $J_j \equiv J$ without lack of generality. The translational symmetry of circulant matrices, including the linear coupling matrix $\mathrm{J}$, allows for the analytical solution of periodic systems. For more on the general properties of circulant matrices, see \cite{gray2006toeplitz}. Leveraging these properties, the circulant matrix $\mathrm{J}$} can be easily diagonalized using the discrete Fourier matrix $S$, a complex symmetric and unitary matrix with elements \cite{hudgings2002design}
\begin{equation}\label{Fourier}
   S_{j,p}=\frac{1}{\sqrt{N}} e^{i\frac{ 2 \pi j p}{N}}.
\end{equation}
The elements of the Fourier matrix fulfill the following orthonormality conditions 
\begin{equation*}
   \sum_{j=1}^{N} S_{j,p} S^{*}_{j,q}=\delta_{p,q}, \qquad \sum_{p=1}^{N} S_{j,p} S^{*}_{k,p}=\delta_{j,k}.
\end{equation*}
The \textit{N} eigenvectors of the coupling matrix $C$ are the linear supermodes of the system $\mathcal{B}$ and form a basis. From now on, we refer to them as Fourier modes as they are the columns of the discrete Fourier matrix $S$. Fourier and individual mode bases are thus related by 
\begin{equation*}
    \hat{\mathcal{A}}_{j}=\sum_{p=1}^{N} S_{j,p}\,\hat{\mathcal{B}}_{p}, \qquad \hat{\mathcal{B}}_{p}=\sum_{j=1}^{N} S^{*}_{j,p}\,\hat{\mathcal{A}}_{j},
\end{equation*}
\DBR{which helps to decouple the linear dynamics of the waveguide array, allowing for a closed-form analytical treatment of the system.} The \textit{N} eigenvalues of $C$ are the propagation constants of the Fourier modes, given by
\begin{align}
\label{four}
   &\lambda_{p} = 2 J \cos{\left (\frac{2 \pi p}{N}\right)} =\lambda_{N-p}, 
   &\lambda_{p} =-\lambda_{\frac{N}{2}-p}, 
\end{align} 
where $p = 1,...,N$.

In a circular array without nonlinearity, if we excite the $p$th Fourier mode at the input of the array, it will propagate without changing its shape with propagation constant $\lambda_{p}$. Notably, arrays with \DBR{$N=4n$ $(n = 1,2,...)$ number of waveguides} have two Fourier modes with $\lambda_{p}=0$.

Particularly, the Fourier modes of a circular array of identical waveguides fulfill
\begin{align*}
    &S^{*}_{j,N-p}=S_{j,p},\\
    &S^{*}_{j,\frac{N}{2}-p}=(-1)^{j} S_{j,p}  ~~\text{(N even)},\\
    &S^{*}_{j,\frac{3N}{4}-p}=i^{j} S_{j,p}  ~~\text{(N multiple of 4)} .
\end{align*} 
These properties result in the following useful orthonormality relations

\begin{align}
\label{five}
   &\sum_{j=1}^{N} S_{j,p} S_{j,q} =\delta_{p,N-q}. \\ 
   &\sum_{j=1}^{N} (-1)^{j} S_{j,p} S_{j,q} = \delta_{p,\frac{N}{2}-q} ~~\text{(N even)}.\\
   &\sum_{j=1}^{N} i^{j} S_{j,p} S_{j,q} = \delta_{p,\frac{3N}{4}-q} ~~\text{(N multiple of 4)}.
\end{align} 
Moreover, Fourier modes fulfill the following interesting property
\begin{equation}\label{Gold}
   \sum_{j=1}^{N} S_{j,r} S_{j,p} S_{j,q} = \frac{1}{\sqrt{N}} \sum_{j=1}^{N} S_{j,p} S_{j,q+r}=\frac{\delta_{p,N-(q+r)}}{\sqrt{N}}. 
\end{equation} 
Thus, Equations (6)-(8) are special cases of this relation for $r=0, N/2$, and $N/4$, up to a constant factor $\sqrt{N}$. 

In the following section, we show how to obtain analytical solutions when pumping the circular ANW with a pump with amplitude and phase profiles proportional to the coefficients $S_{j,r}$.

\section{\label{IV} Analytical solutions for the circular array of nonlinear waveguides}

Fourier modes \eqref{Fourier} and the relation \eqref{Gold} \SM{enable us} to solve analytically the system of Equations \eqref{two} for specific pump profiles. Rewriting Equation \eqref{two} in terms of Fourier modes we obtain
 \begin{equation*}
\frac{d {\hat{\mathcal{B}}}_{p}}{d z}=-i\lambda_{p}\hat{\mathcal{B}}_{p}-2i\sum_{j=1}^{N} \sum_{q=1}^{N} \eta_{j} S^{*}_{j,p}\,S^{*}_{j,q} \hat{\mathcal{B}}_{q}^{\dagger}.
\end{equation*}
Interestingly, pumping the array with a profile following $\eta_{j} =\vert \eta \vert S^{*}_{j,r}$ and using the complex conjugate of relation \eqref{Gold}, we obtain
 \begin{equation}\label{GenCase}
\frac{d {\hat{\mathcal{B}}}_{p}}{d z}=-i\lambda_{p}\hat{\mathcal{B}}_{p}-2i\vert \eta  \vert \hat{\mathcal{B}}_{N-(p+r)}^{\dagger},
\end{equation}
where we have absorbed the factor $N^{-1/2}$ in $\vert\eta\vert$. This is the equation of a two-mode squeezer coupling modes $p$ and $(N-r)-p$ and has analytical solutions for any number of waveguides \textit{N}. 

The general solution of Equation (\ref{GenCase}) can be written as
\begin{align} \label{r-sol}
\hat{\mathcal{B}}_{p}(z)&=[\cos(F_{p} z) \hat{\mathcal{B}}_{p}(0) -i\frac{\sin(F_{p} z) }{2 F_{p}} \{(\lambda_{p} + \lambda_{N-(p+r)})\times   \nonumber\\&\hat{\mathcal{B}}_{p}(0)+ 4 \vert \eta \vert \,\hat{\mathcal{B}}_{N-(p+r)}^{\dagger}(0)\}]e^{- \frac{i}{2} (\lambda_{p} - \lambda_{N-(p+r)}) z},
\end{align}
where $F_{p} = \sqrt{(\lambda_{p}+\lambda_{N-(p+r)})^2 - 16\vert\eta\vert^2}/2 $. Remarkably, Equation (\ref{r-sol}) is valid for any number of waveguides $N$ and can be calculated at any propagation distance $z$ for any coupling profile $J_j$: homogeneous, parabolic, Glauber-Fock, etc \cite{barral2020quantum}. 

In the next subsections, \SM{we show particular solutions for the cases $r=0$ (uniform pump), $r=N/2$ (alternating $\pi$-phase pump), and $r=N/4$ (alternating $\pi/2$-phase pump)}.

\subsection{Pump profile uniform in amplitude and phase (r=0)}
In this case we obtain from Equation \eqref{r-sol} the following exact solution
\begin{align} \label{p-sol}
\hat{\mathcal{B}}_{p}(z)=\cos(F_{p} z) \hat{\mathcal{B}}_{p}(0) -i\frac{\sin(F_{p} z) }{F_{p}}[\lambda_{p} \hat{\mathcal{B}}_{p}(0) ~~~~~~~~\nonumber \\ + 2 \vert \eta \vert \,\hat{\mathcal{B}}_{N-p}^{\dagger}(0)],
\end{align}
where $F_{p} = \sqrt{\lambda_{p}^2 - 4 \vert\eta\vert^2}$ and we have used $\lambda_p = \lambda_{N-p}$ from Equation (\ref{four}). The value of $F_p$ defines two operation regimes under standard pump power and coupling conditions: i) $\vert \lambda_{p} \vert > 2 \vert \eta \vert$, where linear coupling dominates over nonlinear interactions, promoting spreading of SPDC light for multimode entanglement generation; and ii) $\vert \lambda_{p} \vert < 2 \vert \eta \vert$, where confined SPDC renders the array as decoupled single-mode squeezers \cite{fiuravsek2000substituting}. Therefore we focus on the entanglement-favorable power regime of  $\vert \lambda_{p} \vert > 2 \vert \eta \vert$.

Notably, the solution of Equation \eqref{p-sol} for Fourier modes with zero eigenvalues ($\lambda_{p} = 0$) is hyperbolic and given by
\begin{align} \label{z-sol}
\hat{\mathcal{B}}_{p}(z)=\cosh(2 \vert \eta \vert z) \hat{\mathcal{B}}_{p}(0) -i\sinh(2 \vert \eta \vert z) \hat{\mathcal{B}}_{N-p}^{\dagger}(0),
\end{align}
where $p \equiv l, l^{'}$ corresponds to the zero Fourier modes. These correspond to the Fourier modes that fulfill $p=(2n+1)N/4$ with $n=0,1$ and its shape is $S_{j,p}=(\pm i)^{j}/\sqrt{N}$ with $\lambda_{l}=\lambda_{l'}= 0$. Note that Equations (\ref{p-sol}) and (\ref{z-sol}) represent the solutions of non-phase-matched and perfectly phase-matched Fourier modes, respectively \cite{mollow1967quantum}. Remarkably, the zero Fourier modes are efficiently built up through SPDC at all propagation distances as they are perfectly phase-matched and, interestingly, for high coupling both zero Fourier modes quickly dominate over other Fourier modes as these are the only Fourier modes that remain phase-matched throughout the propagation length. 

Back to the individual mode basis, the Fourier-mode solution of Equation (\ref{p-sol}) can be written as
\begin{equation}\label{IndSol1}
\mathcal{\hat{A}}_{j}(z) = \sum_{j'=1}^{N} [\mathbf{U}_{j, j'}(z) \mathcal{\hat{A}}_{j'}(0)-\mathbf{U}^{'}_{j, j'}(z) \mathcal{\hat{A}}_{j'}^{\dagger}(0) ],
\end{equation}
where
\begin{align} \nonumber
\mathbf{U}_{j,j'}(z)=\sum_{p=1}^{N} &S_{j,p} S^{*}_{p, j'} \left[\cos(F_{p} z)-i \frac{\lambda_{p}}{F_{p}} \sin(F_{p} z)\right], \\  \nonumber
\mathbf{U}^{'}_{j, j'}(z)=\sum_{p=1}^{N} &S_{j,p} S^{*}_{p, j'} \left[\frac{2 i \vert\eta\vert} {F_{p}} \sin(F_{p} z)\right],        
\end{align}
with 
\begin{align} \nonumber
\sum_{j=1}^{N} \left[\vert \mathbf{U}_{j,j'}(z) \vert^2 - \vert \mathbf{U}^{'}_{j, j'}(z) \vert^2\right] = 1.       
\end{align}
Note that for $\vert \eta \vert = 0$, $\mathbf{U}_{j,j'}(z) = \mathbf{\tilde{U}}_{j,j'}(z)$ and $\mathbf{U}^{'}_{j, j'}(z) = 0$, with $\mathbf{\tilde{U}}_{j,j'}(z) = \sum_{p=1}^{N} S_{j,p} S^{*}_{p, j'}e^{-i \lambda_{p}z} $. Equations (\ref{p-sol}) and (\ref{IndSol1}) are analytical solutions to the propagation problem in circular ANWs in the Fourier and the individual mode bases, respectively. These two solutions serve as a resource for encoding quantum information in two different bases.

The relevant observables to analyze continuous-variable entanglement are the quadratures of the electric field. Thus, we write Equation (\ref{p-sol}) in terms of Fourier-mode quadratures as 
\begin{equation}
\label{nsmfp}
    \hat{\xi}_{\mathcal{S},p}(z) = \mathcal{S}_{p} (z)  \hat{\mathcal{\xi}}_{\mathcal{S},p}(0),
\end{equation}
with
\begin{equation*}
    \mathcal{S}_{p}(z) = \begin{bmatrix}
\cos{(F_{p}z)} & \pm e^{-r_{p}} \sin{(F_{p}z)} \\
\mp e^{r_{p}} \sin{(F_{p}z)} & \cos{(F_{p}z)}
\end{bmatrix},
\end{equation*}
where $r_{p}=(1/2) \ln{[(\lambda_{p}+2\vert \eta \vert)/(\lambda_{p}-2\vert \eta \vert)]}$ and the signs $\pm(\mp)$ apply to the modes $N,...,\frac{N}{2}+1$ and $\frac{N}{2},...,1$, respectively.

As the quantum states produced through SPDC are Gaussian and with zero mean, the second-order moments of the quadrature operators gather all the information about the quantum state, and we can arrange them in the form of a covariance matrix ${V(z)}$ \cite{adesso2014continuous}. When the input is a vacuum state, the elements of the covariance matrix $V(z)$ can be obtained from Equation (\ref{nsmfp}) using $V(z)=\mathcal{S}(z) V(0) \mathcal{S}^{T}(z)$ with $V(0)$ a $2N\times 2N$ identity matrix and $\mathcal{S}(z)$ the propagator in a given basis --individual or Fourier. From the elements of this matrix, quantum correlations and the evolution of variances for Fourier modes can be obtained at any propagation length $z$. 

\begin{figure*}[t]
{\raisebox{0.2cm}{ \hspace{0.2cm}}}
\stackunder[8pt]{${\bf  r = 0}$}{\subfigure{\includegraphics[width=0.313\textwidth]{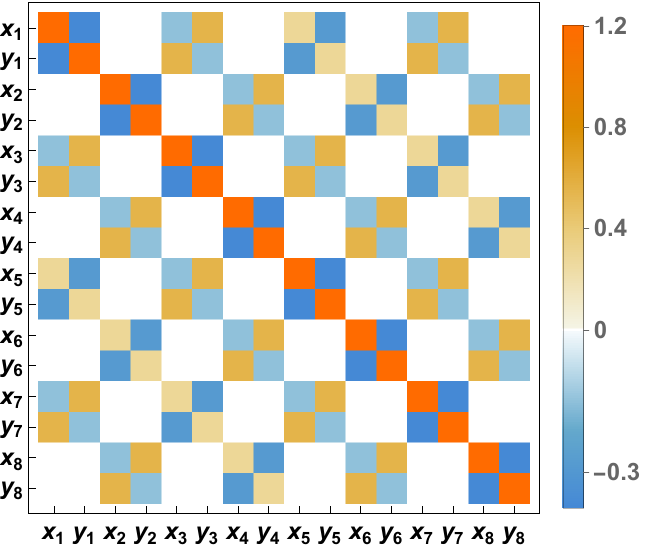}}}  \put(-41,-22){(a)} \hspace{0.1cm}
\stackunder[8pt] {${\bf r = N/2}$}{\subfigure{\includegraphics[width=0.313\textwidth]{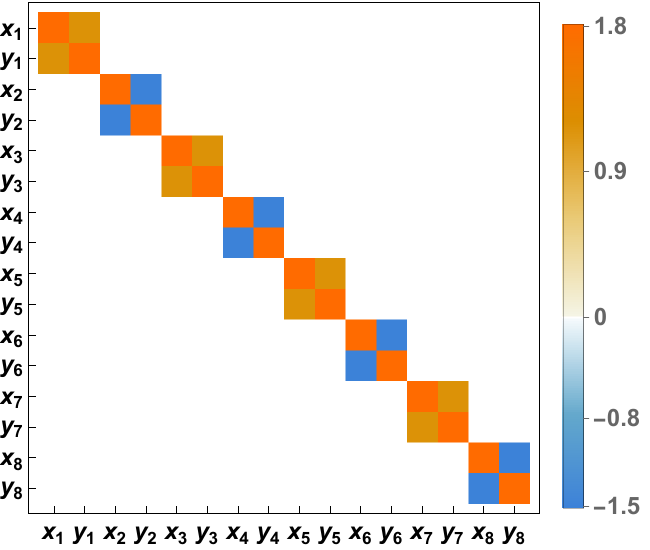}}} \put(-41,-24){(b)}\hspace{0.1cm}
\stackunder[8pt]{${\bf  r = N/4}$}{\subfigure{\includegraphics[width=0.313\textwidth]{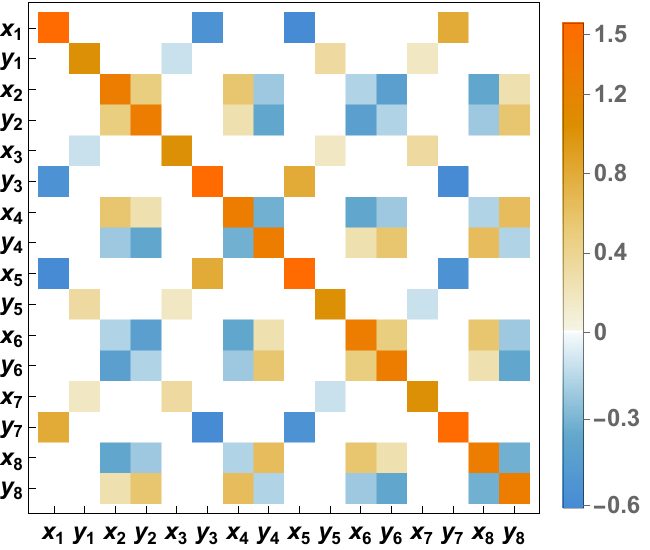}}} \put(-41,-24){(c)}\hspace{0.1cm}
\caption{\label{F2}\small{Covariance matrices V(z) in the individual mode basis for an eight-waveguide circular array with homogeneous coupling profile. (a) Quantum correlations for a flat pump profile with uniform amplitude and uniform phase (r = 0). For a flat pump profile with an alternative $\pi$ phase, we obtained a block diagonal matrix as displayed in (b). Quantum correlations obtained for a flat pump profile with an alternative $\pi/2$ phase are shown in (c). We chopped absolute values lower than $10^{-2}$ from covariance matrices for exposition. The coupling and nonlinearity parameters as follows: J = 0.45 $\text{mm}^{-1}$, $\eta$ = 0.015 $\text{mm}^{-1}$, and z = 20 mm. }}  
\end{figure*}

A simple change of basis is sufficient to get the covariance matrix in the individual-mode basis. In terms of individual-quadratures, the elements of the covariance matrix $V(z)$ are
\begin{align}
   & V(x_{i}, x_{j}) = \sum_{p=1}^{N}[S_{i,p} S^{*}_{p, j} (\zeta -\beta -\gamma  ) +  S^{*}_{i,p} S_{p, j} ( \beta -\gamma+\delta) ],
  \nonumber  \\ 
     &V(y_{i}, y_{j}) = \sum_{p=1}^{N} [S_{i,p} S^{*}_{p, j} (\zeta +\beta +\gamma  )  +  S^{*}_{i,p} S_{p, j} ( \gamma-\beta + \delta)],
     \nonumber  \\ \label{COV}
      &V(x_{i}, y_{j}) = \sum_{p=1}^{N}i [S_{i,p} S^{*}_{p, j}  (\beta +\gamma)   +  S^{*}_{i,p} S_{p, j}  (\beta - \gamma) ], 
\end{align}
where
\begin{align*}
    &\zeta = \cos^{2}(F_{p} z)+ \frac{\lambda_{p}^2}{F_{p}^2} \sin^{2}(F_{p} z), \nonumber  \\ &\beta = \frac{ 2 i \vert\eta\vert}{F_{p}}  \sin{(F_{p}z)}  \cos{(F_{p}z)}, \nonumber  \\ &\gamma = \frac{ 2 \vert\eta\vert \lambda_{p}}{F_{p}^2}\sin^{2}{(F_{p}z)},~ \quad \delta = \frac{ 4 \vert\eta\vert^{2} }{F_{p}^2}\sin^{2}{(F_{p}z),}
\end{align*}
The presence of off-diagonal elements in the covariance matrix suggests that the circular array with a uniform pump profile in amplitude and phase generates quantum correlations between the individual modes, as illustrated in Fig. \ref{F2}(a) -- making entanglement possible in that basis. Interestingly, the results presented in this section are based on the orthogonality relations of Fourier modes and are general for any circular ANWs with any evanescent coupling profile $J_j$. We will use this covariance matrix in Section V to analyze the full inseparability of the generated quantum state with this pump profile. 

\subsection{Pump profile with uniform amplitude, varying phase (r=N/2)}

When all the waveguides in an N-waveguide circular array are equally pumped such that $\eta_{j} = (-1)^{j}~ \vert \eta \vert $, i.e., using an alternating $\pi$ phase, we obtain the following exact solution using Equation (11) as

\begin{align} \label{vp-sol}
\hat{\mathcal{B}}_{p}(z)=[\cosh(2 \vert \eta \vert z) \hat{\mathcal{B}}_{p}(0) -i\sinh(2 \vert \eta \vert z) \hat{\mathcal{B}}_{\frac{N}{2}-p}^{\dagger}(0)] e^{-i \lambda_{p}z}.
\end{align}
The Fourier-mode solution of Equation (\ref{vp-sol}) in the individual mode basis can be written as

\begin{align}\label{IndSol2}
\mathcal{\hat{A}}_{j}(z) = \sum_{j'=1}^{N} \mathbf{\tilde{U}}_{j,j'}(z)[\cosh{(2\vert\eta\vert z)} \mathcal{\hat{A}}_{j'}(0) ~~~~~~~~~~~\nonumber \\ -(-1)^{j'} i  \sinh{(2\vert\eta\vert z)} \mathcal{\hat{A}}_{j'}^{\dagger}(0) ],
\end{align}
where we have used Equation (\ref{four}), properties of the Fourier matrix and $\mathbf{\tilde{U}}_{j,j'}(z)$ introduced above. After a thorough yet straightforward calculation, we derive the elements of the covariance matrix V(z) using Equation (\ref{IndSol2}), which are as follows:
\begin{align}
    &V(x_{i},x_{j}) = \cosh{(4\vert\eta\vert z)} \delta_{i,j} ,\nonumber \\
      &V(y_{i},y_{j}) = \cosh{(4\vert\eta\vert z)} \delta_{i,j} ,\nonumber \\ \label{COV1}
        &V(x_{i},y_{j}) = -(-1)^{j} \sinh{(4\vert\eta\vert z)} \delta_{i,j} .
\end{align}
These elements form a covariance matrix with a block diagonal structure, where each $2 \times 2$ block corresponds to the correlations of each mode. Equation (\ref{COV1}) implies that no off-diagonal components are present, which would produce quantum correlations between distinct modes of the array. The circular array thus generates independent squeezed fields.  As one can see, quantum correlations are efficiently generated in the individual mode basis for the pump profile with uniform amplitude and phase, which disappears for the pump profile with varying phase, as shown in Fig. \ref{F2}(b). \DBR{This setup thus allows for an active on/off switching of entanglement with application in quantum communications.}

\subsection{Pump profile with uniform amplitude, varying phase (r=N/4)}

Another interesting case appears when waveguides of the circular array are equally pumped such that $\eta_{j} = (-i)^{j}~ \vert \eta \vert $, i.e., using an alternating $\pi/2$ phase. We obtain the following exact solution using Equation (11) as

\begin{align} \label{r-sol3}
\hat{\mathcal{B}}_{p}(z)=[\cos(F_{p} z) \hat{\mathcal{B}}_{p}(0) -i\frac{\sin(F_{p} z) }{2 F_{p}} \{(\lambda_{p} + \lambda_{\frac{3N}{4}-p})\times   \nonumber\\ \hat{\mathcal{B}}_{p}(0)+ 4 \vert \eta \vert \,\hat{\mathcal{B}}_{\frac{3N}{4}-p}^{\dagger}(0)\}]e^{- i (\lambda_{p} - \lambda_{\frac{3N}{4}-p}) z/2}.
\end{align}
In the individual mode basis, the solution (\ref{r-sol3}) can be written as
\begin{align}\label{IndSol3}
\mathcal{\hat{A}}_{j}(z) = \sum_{j'=1}^{N} \mathbf{\bar{U}}_{j,j'}(z)[\tilde{\beta} \mathcal{\hat{A}}_{j'}(0) -(-i)^{j'}   \tilde{\delta} \mathcal{\hat{A}}_{j'}^{\dagger}(0) ],
\end{align}
where
\begin{align*}
  &\mathbf{\bar{U}}_{j,j'}(z)=\sum_{p=1}^{N} S_{j,p} S^{*}_{p, j'} e^{-i(\lambda_{p}-\lambda_{\frac{3N}{4}-p})z/2}, \\  \nonumber
    &\tilde{\beta} = \cos{(F_{p}z)}-i\frac{\sin{(F_{p}z)}}{2F_{p}}(\lambda_{p}+\lambda_{\frac{3N}{4}-p}), \nonumber  \\ &\tilde{\delta} = \frac{ 2 i \vert\eta\vert}{F_{p}}  \sin{(F_{p}z)}.
\end{align*}
From the Equation (\ref{IndSol3}), we obtain the elements of the covariance matrix $V(z)$ as:
\begin{align}
    V(x_{i}, x_{j}) = \sum_{p=1}^{N}[S_{i,p} S^{*}_{p, j} \{-(-i)^{j} \tilde{\beta} \tilde{\delta} e^{-i(\lambda_{p}-\lambda_{\frac{3N}{4}-p})z} + \tilde{\beta} \tilde{\beta}^{*}\} + \nonumber\\  S^{*}_{i,p} S_{p, j} \{-(i)^{j} \tilde{\beta}^{*} \tilde{\delta}^{*} e^{i(\lambda_{p}-\lambda_{\frac{3N}{4}-p})z} + \tilde{\delta} \tilde{\delta}^{*}\}] \nonumber,
   \end{align} 
\begin{align}
     V(y_{i}, y_{j}) = \sum_{p=1}^{N}[S_{i,p} S^{*}_{p, j} \{(-i)^{j} \tilde{\beta} \tilde{\delta} e^{-i(\lambda_{p}-\lambda_{\frac{3N}{4}-p})z} + \tilde{\beta} \tilde{\beta}^{*}\} + \nonumber \\  S^{*}_{i,p} S_{p, j} \{(i)^{j} \tilde{\beta}^{*} \tilde{\delta}^{*} e^{i(\lambda_{p}-\lambda_{\frac{3N}{4}-p})z} + \tilde{\delta} \tilde{\delta}^{*}\}] \nonumber,
\end{align}
 \begin{align}
     \label{COV3}
      V(x_{i}, y_{j}) = \sum_{p=1}^{N}[S_{i,p} S^{*}_{p, j} \{(-i)^{j} \tilde{\beta} \tilde{\delta} e^{-i(\lambda_{p}-\lambda_{\frac{3N}{4}-p})z} \} -  ~~~~~~~~~ \nonumber \\ S^{*}_{i,p} S_{p, j} \{(-i)^{j} \tilde{\beta}^{*} \tilde{\delta}^{*} e^{i(\lambda_{p}-\lambda_{\frac{3N}{4}-p})z} \}], 
\end{align}
Here, off-diagonal elements of the covariance matrix are present, as shown in Fig. \ref{F2}(c). Compared with the flat pump Fig. \ref{F2}(a), it produces a different set of quantum correlations between the individual modes as shown in Equations (\ref{COV3}).

\section{\label{V} Multipartite entanglement in circular ANWs}
To detect genuine multipartite CV entanglement, we use experimentally accessible criteria -- known as the van Loock-Furusawa (VLF) inequalities -- based on the variances of specific quadrature combinations that can be measured with simple homodyne detection \cite{van2003detecting, braunstein2005quantum}. Simultaneous violation of a set of VLF inequalities detects complete inseparability and provides a sufficient condition for genuine multimode entanglement in N-mode CV systems. We verify the inseparability for our proposed configuration and demonstrate multipartite entanglement in the individual mode basis through violation of the following VLF criteria
\begin{align}
\text{VLF}_{k}\equiv V\left[{x}_{k}({\theta_{k}}) - {x}_{k+1}(\theta_{k+1})\right] + \nonumber~~~~~~~~~~~~~~~\\  \label{VLF} ~~~ V\left[{y}_{k}(\theta_{k}) + {y}_{k+1}(\theta_{k+1})\right]  \geq 4,
\end{align}
where $\hat{x}_{k}({\theta_{k}})=\hat{x}_{k} \cos{(\theta_{k})}+\hat{y}_{k} \sin{(\theta_{k})}$ and $\hat{y}_{k}({\theta_{k}})=-\hat{x}_{k} \sin{(\theta_{k})}+\hat{y}_{k} \cos{(\theta_{k})}$ are generalized amplitude and phase quadratures with index $k$ in a given basis. The violation depth of VLF inequalities strongly depends on the local oscillator phase $\theta_{k}$ and should be appropriately tuned to detect the full inseparability.

Generally, after numerically solving the propagation in a circular ANWs given by Equation \eqref{two}, we can check the full inseparability of the state using Equation \eqref{VLF}. However, numerical computations do not offer any insight about the entanglement resource. Using our analytical solutions, one can understand the physical process underlying multipartite entanglement. We explain in the following that efficiently-squeezed phase-matched Fourier modes turn out to be a straightforward way of generating multipartite entanglement in circular ANWs with \DBR{$N=4n$} number of waveguides. We select the individual mode basis for measurement, as it is the most accessible basis and allows entanglement distribution in a network. 

\begin{figure}[t]
  \centering  
    \includegraphics[width=0.48\textwidth]{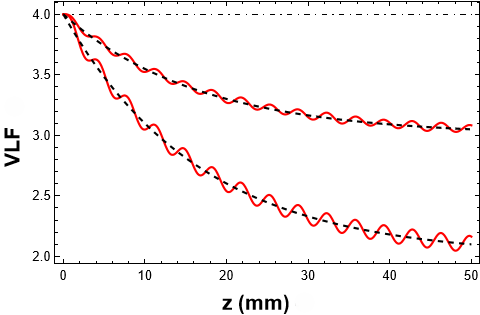}
    \put(-65,52){N = 4}
    \put(-65,107){N = 8}
\caption{\label{F3}\small{\DBR{Multipartite entanglement generation in the individual mode basis versus propagation for four and eight waveguides and a flat pump profile ($\eta_{j} = \vert \eta \vert $). VLF inequalities for a typical value of coupling constant $ J = 0.45 ~ mm^{-1}$ are shown in red. For comparison, VLF inequalities in the unphyisical limit of infinite coupling are shown in dashed black. Simultaneous values under the threshold value VLF = 4 (dot dashed), imply CV multipartite entanglement. $\eta = 0.015 ~ mm^{-1} $.}}}
\end{figure}

We analyze the multipartite entanglement generated with a uniform pump profile (r=0) using the covariance matrix of Equations \eqref{COV}. \DBR{For the sake of understanding, we first examine the infinite coupling limit}. This limit, although not physically realistic since next-to-nearest neighbor effects would appear, provides insight into the dynamics of the system for arrays with $4n$ waveguides, as the zero-Fourier modes become the dominant Fourier modes generated in the circular array. In the limit of large coupling, just the two zero-Fourier modes $S_{j,p}=(\pm i)^{j}/\sqrt{N}$ will be efficiently squeezed as they are phase-matched all along the propagation length. In contrast, the others will suffer a significant mismatch, which will hinder their ability to build up. This generates two sets of individual modes from the sum and difference of zero-Fourier modes, one with the odd individual modes $(i^{j}-(-i)^{j})/\sqrt{N}\propto\sin(j\pi/2)$ and other with even individual modes $(i^{j}+(-i)^{j})/\sqrt{N}\propto\cos(j\pi/2)$. Figure \ref{F3} shows the violation of VLF inequalities for waveguide arrays with $N = 4$ and $N = 8$ propagating modes. \DBR{We use in our simulations values of parameters typically found in ion-exchanged PPLN waveguides \cite{barral2021supermode}: a typical value of coupling constant ($J = 0.45 ~ mm^{-1}$) (red) and the infinite coupling limit ($J = 100 ~ mm^{-1}$ in our simulations) (black dashed).} For $N=4$ we have two sets of entangled modes $1-3$ and $2-4$, while for $N=8$ we got $1-3-5-7$ and $2-4-6-8$. We investigated respectively one and three VLF inequalities for each odd-even set for $N=4$ and 8 waveguides. For that, we used generalized quadratures of Eq. (\ref{VLF}) with $\theta_{k} = 0$ and $\theta_{k+1} = \pi/2$ to demonstrate the full inseparability of the $N = 4$ and $N = 8$ waveguide systems. We obtained identical plots for $N=4$ for the VLF inequalities in each odd-even set because of the symmetry present in the circular array. Similarly, for $N=8$, we got identical VLF plots in each odd-even set \DBR{for the typical coupling constant (red) and the infinite coupling limit (black dashed)}. The simultaneous violation of these inequalities ($\text{VLF}_{k} < 4$ in Eq. (\ref{VLF})) ensures full inseparability. Since we are dealing with pure states, the propagating signal modes are genuinely multipartite entangled \cite{van2003detecting, braunstein2005quantum}. Moreover, lower coupling strengths $J$ produce oscillations around the large coupling limit due to the contributions of Fourier modes with index $p\not=l,l^{'}$. This contribution leads to stronger entanglement at certain distances.

\begin{figure}[t]
  \centering  
    \includegraphics[width=0.48\textwidth]{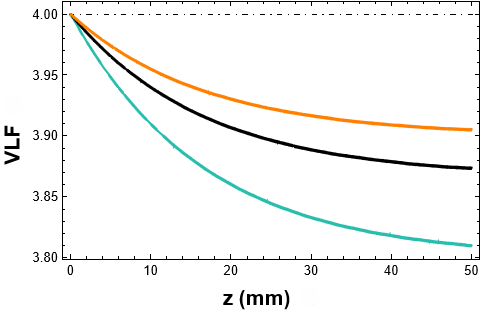}
    \put(-65,46){N = 40}
    \put(-65,82){N = 60}
    \put(-65,103){N = 80}
\caption{\label{F4}\small{VLF inequalities in the limit of \DBR{infinite} coupling for a large number of involved modes: \DBR{$N=40, 60$ and $80$}. Simultaneous values under the threshold value, VLF = 4, imply CV multipartite entanglement. $ J = 100 ~ mm^{-1}.$ $\eta = 0.015 ~ mm^{-1} $.}}
\end{figure}

\DBR{Likewise}, Figure \ref{F4} displays the evolution of multipartite entanglement along propagation for a large number of modes (N = 40, 60, and 80) in the \DBR{infinite} coupling limit. Interestingly, quantum correlations can be observed at any z, regardless of the number of waveguided modes \DBR{$N=4n$} present. The VLF approaches the threshold (VLF = 4) as the number of waveguide modes increases, since when checking even(odd) sets of modes, we do not take into account the odd(even) sets. This is equivalent to having effectively a mixed state that reduces the strength of the quantum correlations and thus the violation of VLF conditions.

\DBR{The influence of losses in entanglement generation can be easily included in our model \cite{leonhardt1997measuring, costas2025spatio}. Considering uniform Gaussian losses the covariance matrix is ${V}_{\eta}=\eta \,V + (1-\eta) {\bf 1}$, with $\eta \in [0,1]$ the effective optical transmittance that takes into account both optical losses and efficiency of detection. Losses couple each mode to a vacuum mode, resulting in an increase of noise that degrades the squeezing and in a decrease of quantum correlations. Applying the covariance matrix ${V}_{\eta}$ in Eq. \eqref{VLF}, we get
\begin{equation}
    \text{VLF}_{k}^{\eta}=\eta \text{VLF}_{k} + 4(1-\eta).
\end{equation}
This equation shows that the full inseparability detected by the VLF inequalities is robust to losses since $\text{VLF}_{k}^{\eta} \rightarrow 4^{-}$ as $\eta\rightarrow 0$, and thus a quantum state where $\text{VLF}_{k}<4$ will remain entangled for any amount of losses but full absorption.}

\DBR{Remarkably, thanks to our analytical framework we have found the pump configuration and detection basis required to produce and measure a robust multipartite full inseparability across arbitrary propagation distances and for any number of waveguides $N=4n$, a regime elusive to numerical methods without insight on the selection of pump profile and detection basis, where an oscillatory behaviour --periodically transitioning between entangled and separable states, heavily reliant on the number of waveguides and the propagation length, has been recently reported \cite{meena2025theoretical}.}

\section{\label{VII}FEASIBILITY AND OUTLOOK\protect}

Finally, we discuss the experimental feasibility of the proposed device using current technology. Lithium niobate stands out as a prime candidate, as it is already a standard for fabricating evanescently coupled waveguide arrays \cite{heinrich2008evanescent}. While 2D geometries have been theoretically explored for applications such as soliton routing \cite{christodoulides2001blocking,szameit2010discrete, meany2015laser}, recent breakthroughs in three-dimensional waveguide integration have overcome previous fabrication hurdles \cite{li2022femtosecond, xu2022femtosecond}. These advances now provide the necessary technical maturity to physically realize the monolithic circular architecture proposed here. Moreover, these methods can be extended to other nonlinear material platforms, including lithium tantalate and potassium titanyl phosphate. Periodically poled waveguides in lithium niobate with pump and detection setups are suitable for integration on a single chip \cite{alibart2016quantum, jin2014chip, mondain2019chip}. The specified values of waveguide parameters $\eta$ and \textit{J} meet the device's operational needs and are fully compatible with state-of-the-art fabrication processes. The propagation losses in lithium niobate waveguides are very low -- 0.3 to 0.6 dB/cm in femtosecond laser-written waveguides \cite{osellame2007femtosecond, lu2023femtosecond}, with minimal impact on squeezing and entanglement for typical $cm$ lengths \cite{lenzini2018integrated, mondain2019chip, barral2019coupling}. Our configuration is inherently interferometrically stable and scalable, thereby providing robustness against noise, which makes entanglement generation more feasible in ANWs \cite{mondain2019chip, barral2020versatile}. Remarkably, our framework can be applied to analyze topological protection of CV resources, such as squeezing or entanglement \cite{duenas2021quadrature}. Although we focused on analytical solutions and multipartite CV entanglement, our findings can be applied to study the effect of disorder \cite{kokkinakis2024anderson} and the role of next-to-nearest couplings in ANWs \cite{ren2022topologically, zecchetto2025topological}.

\section{\label{VIII} CONCLUSIONS\protect}
 We have found analytical solutions for second-order $\chi^{(2)}$ nonlinear circular arrays with any number of waveguides \DBR{$N=4n$} in the SPDC regime for the flat pump profile and pump profiles with uniform amplitude and varying phases. We have, in particular, introduced a general framework using orthonormality conditions of the Fourier modes to analyze the generation of multipartite entangled states in nonlinear circular arrays. \DBR{Our waveguide system introduces two zero Fourier modes due to its circular geometry. The efficient buildup of these zero modes that remain perfectly phase-matched during propagation leads to efficient generation of quantum light, making our device an efficient source of CV multipartite entanglement in the experimentally accessible individual-mode basis.  Circular ANWs double the number of entangled modes compared to planar arrays and generate two sets of entangled states, one with even modes and the other with odd modes. The two entanglement \emph{layers} --large spatially entangled states of odd modes and even modes-- behave like parallel, minimally cross-talking quantum channels on the same physical device. Thus, circular ANWs could enable richer, more flexible protocols for quantum networks and measurement-based quantum computing than single-set entanglement (e.g., odd-indexed entangled modes in the planar ANWs) \cite{costas2025spatio}. Our work suggests that the further optimization of waveguide system geometries for more number of zero modes, which could lead to a larger number of entangled modes in the system.} Scalability and stability make our configuration promising with respect to bulk-optics-based approaches. We have demonstrated the versatility of the device, which is not dependent on the specific values of coupling and nonlinearity parameters and can be realized with current technology. Finally, our results suggest that nonlinear circular arrays are a compact source of multipartite entangled states and could be useful for a variety of photonic quantum technologies, including quantum computation \cite{hu2023progress, walmsley2023light}, networking \cite{luo2023recent}, and metrology \cite{barbieri2022optical}.
\\

\section*{Acknowledgements}
D.B. was supported by MICINN through the European Union NextGenerationEU recovery plan (PRTR-C17.I1) and the Galician Regional Government through “Planes Complementarios de I+D+I con las Comunidades Autónomas” in Quantum Communication, and by grant RYC2024-048797-I funded by MICIU/AEI/10.13039/501100011033 and by ESF+.

\bibliographystyle{apsrev4-2}
\bibliography{apssamp}

\begin{thebibliography}{79}%
\makeatletter
\providecommand \@ifxundefined [1]{%
 \@ifx{#1\undefined}
}%
\providecommand \@ifnum [1]{%
 \ifnum #1\expandafter \@firstoftwo
 \else \expandafter \@secondoftwo
 \fi
}%
\providecommand \@ifx [1]{%
 \ifx #1\expandafter \@firstoftwo
 \else \expandafter \@secondoftwo
 \fi
}%
\providecommand \natexlab [1]{#1}%
\providecommand \enquote  [1]{``#1''}%
\providecommand \bibnamefont  [1]{#1}%
\providecommand \bibfnamefont [1]{#1}%
\providecommand \citenamefont [1]{#1}%
\providecommand \href@noop [0]{\@secondoftwo}%
\providecommand \href [0]{\begingroup \@sanitize@url \@href}%
\providecommand \@href[1]{\@@startlink{#1}\@@href}%
\providecommand \@@href[1]{\endgroup#1\@@endlink}%
\providecommand \@sanitize@url [0]{\catcode `\\12\catcode `\$12\catcode `\&12\catcode `\#12\catcode `\^12\catcode `\_12\catcode `\%12\relax}%
\providecommand \@@startlink[1]{}%
\providecommand \@@endlink[0]{}%
\providecommand \url  [0]{\begingroup\@sanitize@url \@url }%
\providecommand \@url [1]{\endgroup\@href {#1}{\urlprefix }}%
\providecommand \urlprefix  [0]{URL }%
\providecommand \Eprint [0]{\href }%
\providecommand \doibase [0]{http://dx.doi.org/}%
\providecommand \selectlanguage [0]{\@gobble}%
\providecommand \bibinfo  [0]{\@secondoftwo}%
\providecommand \bibfield  [0]{\@secondoftwo}%
\providecommand \translation [1]{[#1]}%
\providecommand \BibitemOpen [0]{}%
\providecommand \bibitemStop [0]{}%
\providecommand \bibitemNoStop [0]{.\EOS\space}%
\providecommand \EOS [0]{\spacefactor3000\relax}%
\providecommand \BibitemShut  [1]{\csname bibitem#1\endcsname}%
\let\auto@bib@innerbib\@empty
%</preamble>
\bibitem [{\citenamefont {Cacciapuoti}\ \emph {et~al.}(2024)\citenamefont {Cacciapuoti}, \citenamefont {Illiano}, \citenamefont {Viscardi},\ and\ \citenamefont {Caleffi}}]{cacciapuoti2024multipartite}%
  \BibitemOpen
  \bibfield  {author} {\bibinfo {author} {\bibfnamefont {A.~S.}\ \bibnamefont {Cacciapuoti}}, \bibinfo {author} {\bibfnamefont {J.}~\bibnamefont {Illiano}}, \bibinfo {author} {\bibfnamefont {M.}~\bibnamefont {Viscardi}}, \ and\ \bibinfo {author} {\bibfnamefont {M.}~\bibnamefont {Caleffi}},\ }\href@noop {} {\bibfield  {journal} {\bibinfo  {journal} {IEEE Transactions on Network and Service Management}\ } (\bibinfo {year} {2024})}\BibitemShut {NoStop}%
\bibitem [{\citenamefont {Zhuang}\ \emph {et~al.}(2018)\citenamefont {Zhuang}, \citenamefont {Zhang},\ and\ \citenamefont {Shapiro}}]{zhuang2018distributed}%
  \BibitemOpen
  \bibfield  {author} {\bibinfo {author} {\bibfnamefont {Q.}~\bibnamefont {Zhuang}}, \bibinfo {author} {\bibfnamefont {Z.}~\bibnamefont {Zhang}}, \ and\ \bibinfo {author} {\bibfnamefont {J.~H.}\ \bibnamefont {Shapiro}},\ }\href@noop {} {\bibfield  {journal} {\bibinfo  {journal} {Physical Review A}\ }\textbf {\bibinfo {volume} {97}},\ \bibinfo {pages} {032329} (\bibinfo {year} {2018})}\BibitemShut {NoStop}%
\bibitem [{\citenamefont {Santra}\ \emph {et~al.}(2025)\citenamefont {Santra}, \citenamefont {Roy}, \citenamefont {Egger},\ and\ \citenamefont {Hauke}}]{santra2025genuine}%
  \BibitemOpen
  \bibfield  {author} {\bibinfo {author} {\bibfnamefont {G.~C.}\ \bibnamefont {Santra}}, \bibinfo {author} {\bibfnamefont {S.~S.}\ \bibnamefont {Roy}}, \bibinfo {author} {\bibfnamefont {D.~J.}\ \bibnamefont {Egger}}, \ and\ \bibinfo {author} {\bibfnamefont {P.}~\bibnamefont {Hauke}},\ }\href@noop {} {\bibfield  {journal} {\bibinfo  {journal} {Physical Review A}\ }\textbf {\bibinfo {volume} {111}},\ \bibinfo {pages} {022434} (\bibinfo {year} {2025})}\BibitemShut {NoStop}%
\bibitem [{\citenamefont {Yukawa}\ \emph {et~al.}(2008)\citenamefont {Yukawa}, \citenamefont {Ukai}, \citenamefont {Van~Loock},\ and\ \citenamefont {Furusawa}}]{yukawa2008experimental}%
  \BibitemOpen
  \bibfield  {author} {\bibinfo {author} {\bibfnamefont {M.}~\bibnamefont {Yukawa}}, \bibinfo {author} {\bibfnamefont {R.}~\bibnamefont {Ukai}}, \bibinfo {author} {\bibfnamefont {P.}~\bibnamefont {Van~Loock}}, \ and\ \bibinfo {author} {\bibfnamefont {A.}~\bibnamefont {Furusawa}},\ }\href@noop {} {\bibfield  {journal} {\bibinfo  {journal} {Physical Review A—Atomic, Molecular, and Optical Physics}\ }\textbf {\bibinfo {volume} {78}},\ \bibinfo {pages} {012301} (\bibinfo {year} {2008})}\BibitemShut {NoStop}%
\bibitem [{\citenamefont {Cai}\ \emph {et~al.}(2017)\citenamefont {Cai}, \citenamefont {Roslund}, \citenamefont {Ferrini}, \citenamefont {Arzani}, \citenamefont {Xu}, \citenamefont {Fabre},\ and\ \citenamefont {Treps}}]{cai2017multimode}%
  \BibitemOpen
  \bibfield  {author} {\bibinfo {author} {\bibfnamefont {Y.}~\bibnamefont {Cai}}, \bibinfo {author} {\bibfnamefont {J.}~\bibnamefont {Roslund}}, \bibinfo {author} {\bibfnamefont {G.}~\bibnamefont {Ferrini}}, \bibinfo {author} {\bibfnamefont {F.}~\bibnamefont {Arzani}}, \bibinfo {author} {\bibfnamefont {X.}~\bibnamefont {Xu}}, \bibinfo {author} {\bibfnamefont {C.}~\bibnamefont {Fabre}}, \ and\ \bibinfo {author} {\bibfnamefont {N.}~\bibnamefont {Treps}},\ }\href@noop {} {\bibfield  {journal} {\bibinfo  {journal} {Nature communications}\ }\textbf {\bibinfo {volume} {8}},\ \bibinfo {pages} {15645} (\bibinfo {year} {2017})}\BibitemShut {NoStop}%
\bibitem [{\citenamefont {Armstrong}\ \emph {et~al.}(2012)\citenamefont {Armstrong}, \citenamefont {Morizur}, \citenamefont {Janousek}, \citenamefont {Hage}, \citenamefont {Treps}, \citenamefont {Lam},\ and\ \citenamefont {Bachor}}]{armstrong2012programmable}%
  \BibitemOpen
  \bibfield  {author} {\bibinfo {author} {\bibfnamefont {S.}~\bibnamefont {Armstrong}}, \bibinfo {author} {\bibfnamefont {J.-F.}\ \bibnamefont {Morizur}}, \bibinfo {author} {\bibfnamefont {J.}~\bibnamefont {Janousek}}, \bibinfo {author} {\bibfnamefont {B.}~\bibnamefont {Hage}}, \bibinfo {author} {\bibfnamefont {N.}~\bibnamefont {Treps}}, \bibinfo {author} {\bibfnamefont {P.~K.}\ \bibnamefont {Lam}}, \ and\ \bibinfo {author} {\bibfnamefont {H.-A.}\ \bibnamefont {Bachor}},\ }\href@noop {} {\bibfield  {journal} {\bibinfo  {journal} {Nature communications}\ }\textbf {\bibinfo {volume} {3}},\ \bibinfo {pages} {1026} (\bibinfo {year} {2012})}\BibitemShut {NoStop}%
\bibitem [{\citenamefont {Larsen}\ \emph {et~al.}(2019)\citenamefont {Larsen}, \citenamefont {Guo}, \citenamefont {Breum}, \citenamefont {Neergaard-Nielsen},\ and\ \citenamefont {Andersen}}]{larsen2019deterministic}%
  \BibitemOpen
  \bibfield  {author} {\bibinfo {author} {\bibfnamefont {M.~V.}\ \bibnamefont {Larsen}}, \bibinfo {author} {\bibfnamefont {X.}~\bibnamefont {Guo}}, \bibinfo {author} {\bibfnamefont {C.~R.}\ \bibnamefont {Breum}}, \bibinfo {author} {\bibfnamefont {J.~S.}\ \bibnamefont {Neergaard-Nielsen}}, \ and\ \bibinfo {author} {\bibfnamefont {U.~L.}\ \bibnamefont {Andersen}},\ }\href@noop {} {\bibfield  {journal} {\bibinfo  {journal} {Science}\ }\textbf {\bibinfo {volume} {366}},\ \bibinfo {pages} {369} (\bibinfo {year} {2019})}\BibitemShut {NoStop}%
\bibitem [{\citenamefont {Asavanant}\ \emph {et~al.}(2019)\citenamefont {Asavanant}, \citenamefont {Shiozawa}, \citenamefont {Yokoyama}, \citenamefont {Charoensombutamon}, \citenamefont {Emura}, \citenamefont {Alexander}, \citenamefont {Takeda}, \citenamefont {Yoshikawa}, \citenamefont {Menicucci}, \citenamefont {Yonezawa} \emph {et~al.}}]{asavanant2019generation}%
  \BibitemOpen
  \bibfield  {author} {\bibinfo {author} {\bibfnamefont {W.}~\bibnamefont {Asavanant}}, \bibinfo {author} {\bibfnamefont {Y.}~\bibnamefont {Shiozawa}}, \bibinfo {author} {\bibfnamefont {S.}~\bibnamefont {Yokoyama}}, \bibinfo {author} {\bibfnamefont {B.}~\bibnamefont {Charoensombutamon}}, \bibinfo {author} {\bibfnamefont {H.}~\bibnamefont {Emura}}, \bibinfo {author} {\bibfnamefont {R.~N.}\ \bibnamefont {Alexander}}, \bibinfo {author} {\bibfnamefont {S.}~\bibnamefont {Takeda}}, \bibinfo {author} {\bibfnamefont {J.-i.}\ \bibnamefont {Yoshikawa}}, \bibinfo {author} {\bibfnamefont {N.~C.}\ \bibnamefont {Menicucci}}, \bibinfo {author} {\bibfnamefont {H.}~\bibnamefont {Yonezawa}},  \emph {et~al.},\ }\href@noop {} {\bibfield  {journal} {\bibinfo  {journal} {Science}\ }\textbf {\bibinfo {volume} {366}},\ \bibinfo {pages} {373} (\bibinfo {year} {2019})}\BibitemShut {NoStop}%
\bibitem [{\citenamefont {Jia}\ \emph {et~al.}(2025)\citenamefont {Jia}, \citenamefont {Zhai}, \citenamefont {Zhu}, \citenamefont {You}, \citenamefont {Cao}, \citenamefont {Zhang}, \citenamefont {Zheng}, \citenamefont {Fu}, \citenamefont {Mao}, \citenamefont {Dai} \emph {et~al.}}]{jia2025continuous}%
  \BibitemOpen
  \bibfield  {author} {\bibinfo {author} {\bibfnamefont {X.}~\bibnamefont {Jia}}, \bibinfo {author} {\bibfnamefont {C.}~\bibnamefont {Zhai}}, \bibinfo {author} {\bibfnamefont {X.}~\bibnamefont {Zhu}}, \bibinfo {author} {\bibfnamefont {C.}~\bibnamefont {You}}, \bibinfo {author} {\bibfnamefont {Y.}~\bibnamefont {Cao}}, \bibinfo {author} {\bibfnamefont {X.}~\bibnamefont {Zhang}}, \bibinfo {author} {\bibfnamefont {Y.}~\bibnamefont {Zheng}}, \bibinfo {author} {\bibfnamefont {Z.}~\bibnamefont {Fu}}, \bibinfo {author} {\bibfnamefont {J.}~\bibnamefont {Mao}}, \bibinfo {author} {\bibfnamefont {T.}~\bibnamefont {Dai}},  \emph {et~al.},\ }\href@noop {} {\bibfield  {journal} {\bibinfo  {journal} {Nature}\ ,\ \bibinfo {pages} {1}} (\bibinfo {year} {2025})}\BibitemShut {NoStop}%
\bibitem [{\citenamefont {Roh}\ \emph {et~al.}(2025)\citenamefont {Roh}, \citenamefont {Gwak}, \citenamefont {Yoon},\ and\ \citenamefont {Ra}}]{roh2025generation}%
  \BibitemOpen
  \bibfield  {author} {\bibinfo {author} {\bibfnamefont {C.}~\bibnamefont {Roh}}, \bibinfo {author} {\bibfnamefont {G.}~\bibnamefont {Gwak}}, \bibinfo {author} {\bibfnamefont {Y.-D.}\ \bibnamefont {Yoon}}, \ and\ \bibinfo {author} {\bibfnamefont {Y.-S.}\ \bibnamefont {Ra}},\ }\href@noop {} {\bibfield  {journal} {\bibinfo  {journal} {Nature Photonics}\ ,\ \bibinfo {pages} {1}} (\bibinfo {year} {2025})}\BibitemShut {NoStop}%
\bibitem [{\citenamefont {Li}\ \emph {et~al.}(2022{\natexlab{a}})\citenamefont {Li}, \citenamefont {Li}, \citenamefont {Chen}, \citenamefont {Feng}, \citenamefont {Yan}, \citenamefont {Zhang}, \citenamefont {Bao}, \citenamefont {Liu}, \citenamefont {Ren}, \citenamefont {Wang} \emph {et~al.}}]{li2022chip}%
  \BibitemOpen
  \bibfield  {author} {\bibinfo {author} {\bibfnamefont {M.}~\bibnamefont {Li}}, \bibinfo {author} {\bibfnamefont {C.}~\bibnamefont {Li}}, \bibinfo {author} {\bibfnamefont {Y.}~\bibnamefont {Chen}}, \bibinfo {author} {\bibfnamefont {L.-T.}\ \bibnamefont {Feng}}, \bibinfo {author} {\bibfnamefont {L.}~\bibnamefont {Yan}}, \bibinfo {author} {\bibfnamefont {Q.}~\bibnamefont {Zhang}}, \bibinfo {author} {\bibfnamefont {J.}~\bibnamefont {Bao}}, \bibinfo {author} {\bibfnamefont {B.-H.}\ \bibnamefont {Liu}}, \bibinfo {author} {\bibfnamefont {X.-F.}\ \bibnamefont {Ren}}, \bibinfo {author} {\bibfnamefont {J.}~\bibnamefont {Wang}},  \emph {et~al.},\ }\href@noop {} {\bibfield  {journal} {\bibinfo  {journal} {Photonics Research}\ }\textbf {\bibinfo {volume} {10}},\ \bibinfo {pages} {1533} (\bibinfo {year} {2022}{\natexlab{a}})}\BibitemShut {NoStop}%
\bibitem [{\citenamefont {Chen}\ \emph {et~al.}(2021)\citenamefont {Chen}, \citenamefont {Fu}, \citenamefont {Gong},\ and\ \citenamefont {Wang}}]{chen2021quantum}%
  \BibitemOpen
  \bibfield  {author} {\bibinfo {author} {\bibfnamefont {X.}~\bibnamefont {Chen}}, \bibinfo {author} {\bibfnamefont {Z.}~\bibnamefont {Fu}}, \bibinfo {author} {\bibfnamefont {Q.}~\bibnamefont {Gong}}, \ and\ \bibinfo {author} {\bibfnamefont {J.}~\bibnamefont {Wang}},\ }\href@noop {} {\bibfield  {journal} {\bibinfo  {journal} {Advanced Photonics}\ }\textbf {\bibinfo {volume} {3}},\ \bibinfo {pages} {064002} (\bibinfo {year} {2021})}\BibitemShut {NoStop}%
\bibitem [{\citenamefont {Hajomer}\ \emph {et~al.}(2024)\citenamefont {Hajomer}, \citenamefont {Derkach}, \citenamefont {Jain}, \citenamefont {Chin}, \citenamefont {Andersen},\ and\ \citenamefont {Gehring}}]{hajomer2024long}%
  \BibitemOpen
  \bibfield  {author} {\bibinfo {author} {\bibfnamefont {A.~A.}\ \bibnamefont {Hajomer}}, \bibinfo {author} {\bibfnamefont {I.}~\bibnamefont {Derkach}}, \bibinfo {author} {\bibfnamefont {N.}~\bibnamefont {Jain}}, \bibinfo {author} {\bibfnamefont {H.-M.}\ \bibnamefont {Chin}}, \bibinfo {author} {\bibfnamefont {U.~L.}\ \bibnamefont {Andersen}}, \ and\ \bibinfo {author} {\bibfnamefont {T.}~\bibnamefont {Gehring}},\ }\href@noop {} {\bibfield  {journal} {\bibinfo  {journal} {Science Advances}\ }\textbf {\bibinfo {volume} {10}},\ \bibinfo {pages} {eadi9474} (\bibinfo {year} {2024})}\BibitemShut {NoStop}%
\bibitem [{\citenamefont {Larsen}\ \emph {et~al.}(2021)\citenamefont {Larsen}, \citenamefont {Chamberland}, \citenamefont {Noh}, \citenamefont {Neergaard-Nielsen},\ and\ \citenamefont {Andersen}}]{larsen2021fault}%
  \BibitemOpen
  \bibfield  {author} {\bibinfo {author} {\bibfnamefont {M.~V.}\ \bibnamefont {Larsen}}, \bibinfo {author} {\bibfnamefont {C.}~\bibnamefont {Chamberland}}, \bibinfo {author} {\bibfnamefont {K.}~\bibnamefont {Noh}}, \bibinfo {author} {\bibfnamefont {J.~S.}\ \bibnamefont {Neergaard-Nielsen}}, \ and\ \bibinfo {author} {\bibfnamefont {U.~L.}\ \bibnamefont {Andersen}},\ }\href@noop {} {\bibfield  {journal} {\bibinfo  {journal} {PRX Quantum}\ }\textbf {\bibinfo {volume} {2}},\ \bibinfo {pages} {030325} (\bibinfo {year} {2021})}\BibitemShut {NoStop}%
\bibitem [{\citenamefont {Ge}\ and\ \citenamefont {Jacobs}(2025)}]{ge2025heisenberg}%
  \BibitemOpen
  \bibfield  {author} {\bibinfo {author} {\bibfnamefont {W.}~\bibnamefont {Ge}}\ and\ \bibinfo {author} {\bibfnamefont {K.}~\bibnamefont {Jacobs}},\ }\href@noop {} {\bibfield  {journal} {\bibinfo  {journal} {Physical Review Letters}\ }\textbf {\bibinfo {volume} {135}},\ \bibinfo {pages} {100801} (\bibinfo {year} {2025})}\BibitemShut {NoStop}%
\bibitem [{\citenamefont {Kouadou}\ \emph {et~al.}(2025)\citenamefont {Kouadou}, \citenamefont {Gozlan}, \citenamefont {Garcia}, \citenamefont {Polizzi}, \citenamefont {Fainsin}, \citenamefont {Paparelle}, \citenamefont {Celis}, \citenamefont {Oriot}, \citenamefont {Aad}, \citenamefont {Namdar} \emph {et~al.}}]{kouadou2025homodyne}%
  \BibitemOpen
  \bibfield  {author} {\bibinfo {author} {\bibfnamefont {T.}~\bibnamefont {Kouadou}}, \bibinfo {author} {\bibfnamefont {E.}~\bibnamefont {Gozlan}}, \bibinfo {author} {\bibfnamefont {L.}~\bibnamefont {Garcia}}, \bibinfo {author} {\bibfnamefont {D.}~\bibnamefont {Polizzi}}, \bibinfo {author} {\bibfnamefont {D.}~\bibnamefont {Fainsin}}, \bibinfo {author} {\bibfnamefont {I.}~\bibnamefont {Paparelle}}, \bibinfo {author} {\bibfnamefont {R.}~\bibnamefont {Celis}}, \bibinfo {author} {\bibfnamefont {B.}~\bibnamefont {Oriot}}, \bibinfo {author} {\bibfnamefont {A.~A.}\ \bibnamefont {Aad}}, \bibinfo {author} {\bibfnamefont {P.}~\bibnamefont {Namdar}},  \emph {et~al.},\ }\href@noop {} {\bibfield  {journal} {\bibinfo  {journal} {arXiv preprint arXiv:2511.04578}\ } (\bibinfo {year} {2025})}\BibitemShut {NoStop}%
\bibitem [{\citenamefont {Tasker}\ \emph {et~al.}(2021)\citenamefont {Tasker}, \citenamefont {Frazer}, \citenamefont {Ferranti}, \citenamefont {Allen}, \citenamefont {Brunel}, \citenamefont {Tanzilli}, \citenamefont {D’Auria},\ and\ \citenamefont {Matthews}}]{tasker2021silicon}%
  \BibitemOpen
  \bibfield  {author} {\bibinfo {author} {\bibfnamefont {J.~F.}\ \bibnamefont {Tasker}}, \bibinfo {author} {\bibfnamefont {J.}~\bibnamefont {Frazer}}, \bibinfo {author} {\bibfnamefont {G.}~\bibnamefont {Ferranti}}, \bibinfo {author} {\bibfnamefont {E.~J.}\ \bibnamefont {Allen}}, \bibinfo {author} {\bibfnamefont {L.~F.}\ \bibnamefont {Brunel}}, \bibinfo {author} {\bibfnamefont {S.}~\bibnamefont {Tanzilli}}, \bibinfo {author} {\bibfnamefont {V.}~\bibnamefont {D’Auria}}, \ and\ \bibinfo {author} {\bibfnamefont {J.~C.}\ \bibnamefont {Matthews}},\ }\href@noop {} {\bibfield  {journal} {\bibinfo  {journal} {Nature Photonics}\ }\textbf {\bibinfo {volume} {15}},\ \bibinfo {pages} {11} (\bibinfo {year} {2021})}\BibitemShut {NoStop}%
\bibitem [{\citenamefont {Masada}\ \emph {et~al.}(2015)\citenamefont {Masada}, \citenamefont {Miyata}, \citenamefont {Politi}, \citenamefont {Hashimoto}, \citenamefont {O'brien},\ and\ \citenamefont {Furusawa}}]{masada2015continuous}%
  \BibitemOpen
  \bibfield  {author} {\bibinfo {author} {\bibfnamefont {G.}~\bibnamefont {Masada}}, \bibinfo {author} {\bibfnamefont {K.}~\bibnamefont {Miyata}}, \bibinfo {author} {\bibfnamefont {A.}~\bibnamefont {Politi}}, \bibinfo {author} {\bibfnamefont {T.}~\bibnamefont {Hashimoto}}, \bibinfo {author} {\bibfnamefont {J.~L.}\ \bibnamefont {O'brien}}, \ and\ \bibinfo {author} {\bibfnamefont {A.}~\bibnamefont {Furusawa}},\ }\href@noop {} {\bibfield  {journal} {\bibinfo  {journal} {Nature Photonics}\ }\textbf {\bibinfo {volume} {9}},\ \bibinfo {pages} {316} (\bibinfo {year} {2015})}\BibitemShut {NoStop}%
\bibitem [{\citenamefont {Lenzini}\ \emph {et~al.}(2018)\citenamefont {Lenzini}, \citenamefont {Janousek}, \citenamefont {Thearle}, \citenamefont {Villa}, \citenamefont {Haylock}, \citenamefont {Kasture}, \citenamefont {Cui}, \citenamefont {Phan}, \citenamefont {Dao}, \citenamefont {Yonezawa} \emph {et~al.}}]{lenzini2018integrated}%
  \BibitemOpen
  \bibfield  {author} {\bibinfo {author} {\bibfnamefont {F.}~\bibnamefont {Lenzini}}, \bibinfo {author} {\bibfnamefont {J.}~\bibnamefont {Janousek}}, \bibinfo {author} {\bibfnamefont {O.}~\bibnamefont {Thearle}}, \bibinfo {author} {\bibfnamefont {M.}~\bibnamefont {Villa}}, \bibinfo {author} {\bibfnamefont {B.}~\bibnamefont {Haylock}}, \bibinfo {author} {\bibfnamefont {S.}~\bibnamefont {Kasture}}, \bibinfo {author} {\bibfnamefont {L.}~\bibnamefont {Cui}}, \bibinfo {author} {\bibfnamefont {H.-P.}\ \bibnamefont {Phan}}, \bibinfo {author} {\bibfnamefont {D.~V.}\ \bibnamefont {Dao}}, \bibinfo {author} {\bibfnamefont {H.}~\bibnamefont {Yonezawa}},  \emph {et~al.},\ }\href@noop {} {\bibfield  {journal} {\bibinfo  {journal} {Science advances}\ }\textbf {\bibinfo {volume} {4}},\ \bibinfo {pages} {eaat9331} (\bibinfo {year} {2018})}\BibitemShut {NoStop}%
\bibitem [{\citenamefont {Moody}\ \emph {et~al.}(2022)\citenamefont {Moody}, \citenamefont {Sorger}, \citenamefont {Blumenthal}, \citenamefont {Juodawlkis}, \citenamefont {Loh}, \citenamefont {Sorace-Agaskar}, \citenamefont {Jones}, \citenamefont {Balram}, \citenamefont {Matthews}, \citenamefont {Laing} \emph {et~al.}}]{moody20222022}%
  \BibitemOpen
  \bibfield  {author} {\bibinfo {author} {\bibfnamefont {G.}~\bibnamefont {Moody}}, \bibinfo {author} {\bibfnamefont {V.~J.}\ \bibnamefont {Sorger}}, \bibinfo {author} {\bibfnamefont {D.~J.}\ \bibnamefont {Blumenthal}}, \bibinfo {author} {\bibfnamefont {P.~W.}\ \bibnamefont {Juodawlkis}}, \bibinfo {author} {\bibfnamefont {W.}~\bibnamefont {Loh}}, \bibinfo {author} {\bibfnamefont {C.}~\bibnamefont {Sorace-Agaskar}}, \bibinfo {author} {\bibfnamefont {A.~E.}\ \bibnamefont {Jones}}, \bibinfo {author} {\bibfnamefont {K.~C.}\ \bibnamefont {Balram}}, \bibinfo {author} {\bibfnamefont {J.~C.}\ \bibnamefont {Matthews}}, \bibinfo {author} {\bibfnamefont {A.}~\bibnamefont {Laing}},  \emph {et~al.},\ }\href@noop {} {\bibfield  {journal} {\bibinfo  {journal} {Journal of Physics: Photonics}\ }\textbf {\bibinfo {volume} {4}},\ \bibinfo {pages} {012501} (\bibinfo {year} {2022})}\BibitemShut {NoStop}%
\bibitem [{\citenamefont {Pelucchi}\ \emph {et~al.}(2022)\citenamefont {Pelucchi}, \citenamefont {Fagas}, \citenamefont {Aharonovich}, \citenamefont {Englund}, \citenamefont {Figueroa}, \citenamefont {Gong}, \citenamefont {Hannes}, \citenamefont {Liu}, \citenamefont {Lu}, \citenamefont {Matsuda} \emph {et~al.}}]{pelucchi2022potential}%
  \BibitemOpen
  \bibfield  {author} {\bibinfo {author} {\bibfnamefont {E.}~\bibnamefont {Pelucchi}}, \bibinfo {author} {\bibfnamefont {G.}~\bibnamefont {Fagas}}, \bibinfo {author} {\bibfnamefont {I.}~\bibnamefont {Aharonovich}}, \bibinfo {author} {\bibfnamefont {D.}~\bibnamefont {Englund}}, \bibinfo {author} {\bibfnamefont {E.}~\bibnamefont {Figueroa}}, \bibinfo {author} {\bibfnamefont {Q.}~\bibnamefont {Gong}}, \bibinfo {author} {\bibfnamefont {H.}~\bibnamefont {Hannes}}, \bibinfo {author} {\bibfnamefont {J.}~\bibnamefont {Liu}}, \bibinfo {author} {\bibfnamefont {C.-Y.}\ \bibnamefont {Lu}}, \bibinfo {author} {\bibfnamefont {N.}~\bibnamefont {Matsuda}},  \emph {et~al.},\ }\href@noop {} {\bibfield  {journal} {\bibinfo  {journal} {Nature Reviews Physics}\ }\textbf {\bibinfo {volume} {4}},\ \bibinfo {pages} {194} (\bibinfo {year} {2022})}\BibitemShut {NoStop}%
\bibitem [{\citenamefont {Clark}\ \emph {et~al.}(2025)\citenamefont {Clark}, \citenamefont {Puzio}, \citenamefont {Green}, \citenamefont {Pradyumna}, \citenamefont {Trojak}, \citenamefont {Politi},\ and\ \citenamefont {Matthews}}]{clark2025integrated}%
  \BibitemOpen
  \bibfield  {author} {\bibinfo {author} {\bibfnamefont {R.~N.}\ \bibnamefont {Clark}}, \bibinfo {author} {\bibfnamefont {B.}~\bibnamefont {Puzio}}, \bibinfo {author} {\bibfnamefont {O.~M.}\ \bibnamefont {Green}}, \bibinfo {author} {\bibfnamefont {S.~T.}\ \bibnamefont {Pradyumna}}, \bibinfo {author} {\bibfnamefont {O.}~\bibnamefont {Trojak}}, \bibinfo {author} {\bibfnamefont {A.}~\bibnamefont {Politi}}, \ and\ \bibinfo {author} {\bibfnamefont {J.~C.}\ \bibnamefont {Matthews}},\ }\href@noop {} {\bibfield  {journal} {\bibinfo  {journal} {arXiv preprint arXiv:2506.04771}\ } (\bibinfo {year} {2025})}\BibitemShut {NoStop}%
\bibitem [{\citenamefont {Yang}\ \emph {et~al.}(2025)\citenamefont {Yang}, \citenamefont {Youssry},\ and\ \citenamefont {Peruzzo}}]{yang2025programmable}%
  \BibitemOpen
  \bibfield  {author} {\bibinfo {author} {\bibfnamefont {Y.}~\bibnamefont {Yang}}, \bibinfo {author} {\bibfnamefont {A.}~\bibnamefont {Youssry}}, \ and\ \bibinfo {author} {\bibfnamefont {A.}~\bibnamefont {Peruzzo}},\ }\href@noop {} {\bibfield  {journal} {\bibinfo  {journal} {arXiv preprint arXiv:2502.12385}\ } (\bibinfo {year} {2025})}\BibitemShut {NoStop}%
\bibitem [{\citenamefont {Christodoulides}\ \emph {et~al.}(2003)\citenamefont {Christodoulides}, \citenamefont {Lederer},\ and\ \citenamefont {Silberberg}}]{christodoulides2003discretizing}%
  \BibitemOpen
  \bibfield  {author} {\bibinfo {author} {\bibfnamefont {D.~N.}\ \bibnamefont {Christodoulides}}, \bibinfo {author} {\bibfnamefont {F.}~\bibnamefont {Lederer}}, \ and\ \bibinfo {author} {\bibfnamefont {Y.}~\bibnamefont {Silberberg}},\ }\href@noop {} {\bibfield  {journal} {\bibinfo  {journal} {Nature}\ }\textbf {\bibinfo {volume} {424}},\ \bibinfo {pages} {817} (\bibinfo {year} {2003})}\BibitemShut {NoStop}%
\bibitem [{\citenamefont {Iwanow}\ \emph {et~al.}(2004)\citenamefont {Iwanow}, \citenamefont {Schiek}, \citenamefont {Stegeman}, \citenamefont {Pertsch}, \citenamefont {Lederer}, \citenamefont {Min},\ and\ \citenamefont {Sohler}}]{iwanow2004observation}%
  \BibitemOpen
  \bibfield  {author} {\bibinfo {author} {\bibfnamefont {R.}~\bibnamefont {Iwanow}}, \bibinfo {author} {\bibfnamefont {R.}~\bibnamefont {Schiek}}, \bibinfo {author} {\bibfnamefont {G.}~\bibnamefont {Stegeman}}, \bibinfo {author} {\bibfnamefont {T.}~\bibnamefont {Pertsch}}, \bibinfo {author} {\bibfnamefont {F.}~\bibnamefont {Lederer}}, \bibinfo {author} {\bibfnamefont {Y.}~\bibnamefont {Min}}, \ and\ \bibinfo {author} {\bibfnamefont {W.}~\bibnamefont {Sohler}},\ }\href@noop {} {\bibfield  {journal} {\bibinfo  {journal} {Physical review letters}\ }\textbf {\bibinfo {volume} {93}},\ \bibinfo {pages} {113902} (\bibinfo {year} {2004})}\BibitemShut {NoStop}%
\bibitem [{\citenamefont {Setzpfandt}\ \emph {et~al.}(2009)\citenamefont {Setzpfandt}, \citenamefont {Neshev}, \citenamefont {Schiek}, \citenamefont {Lederer}, \citenamefont {T{\"u}nnermann},\ and\ \citenamefont {Pertsch}}]{setzpfandt2009competing}%
  \BibitemOpen
  \bibfield  {author} {\bibinfo {author} {\bibfnamefont {F.}~\bibnamefont {Setzpfandt}}, \bibinfo {author} {\bibfnamefont {D.~N.}\ \bibnamefont {Neshev}}, \bibinfo {author} {\bibfnamefont {R.}~\bibnamefont {Schiek}}, \bibinfo {author} {\bibfnamefont {F.}~\bibnamefont {Lederer}}, \bibinfo {author} {\bibfnamefont {A.}~\bibnamefont {T{\"u}nnermann}}, \ and\ \bibinfo {author} {\bibfnamefont {T.}~\bibnamefont {Pertsch}},\ }\href@noop {} {\bibfield  {journal} {\bibinfo  {journal} {Optics letters}\ }\textbf {\bibinfo {volume} {34}},\ \bibinfo {pages} {3589} (\bibinfo {year} {2009})}\BibitemShut {NoStop}%
\bibitem [{\citenamefont {Solntsev}\ \emph {et~al.}(2012)\citenamefont {Solntsev}, \citenamefont {Sukhorukov}, \citenamefont {Neshev},\ and\ \citenamefont {Kivshar}}]{solntsev2012spontaneous}%
  \BibitemOpen
  \bibfield  {author} {\bibinfo {author} {\bibfnamefont {A.~S.}\ \bibnamefont {Solntsev}}, \bibinfo {author} {\bibfnamefont {A.~A.}\ \bibnamefont {Sukhorukov}}, \bibinfo {author} {\bibfnamefont {D.~N.}\ \bibnamefont {Neshev}}, \ and\ \bibinfo {author} {\bibfnamefont {Y.~S.}\ \bibnamefont {Kivshar}},\ }\href@noop {} {\bibfield  {journal} {\bibinfo  {journal} {Physical Review Letters}\ }\textbf {\bibinfo {volume} {108}},\ \bibinfo {pages} {023601} (\bibinfo {year} {2012})}\BibitemShut {NoStop}%
\bibitem [{\citenamefont {Solntsev}\ \emph {et~al.}(2014)\citenamefont {Solntsev}, \citenamefont {Setzpfandt}, \citenamefont {Clark}, \citenamefont {Wu}, \citenamefont {Collins}, \citenamefont {Xiong}, \citenamefont {Schreiber}, \citenamefont {Katzschmann}, \citenamefont {Eilenberger}, \citenamefont {Schiek} \emph {et~al.}}]{solntsev2014generation}%
  \BibitemOpen
  \bibfield  {author} {\bibinfo {author} {\bibfnamefont {A.~S.}\ \bibnamefont {Solntsev}}, \bibinfo {author} {\bibfnamefont {F.}~\bibnamefont {Setzpfandt}}, \bibinfo {author} {\bibfnamefont {A.~S.}\ \bibnamefont {Clark}}, \bibinfo {author} {\bibfnamefont {C.~W.}\ \bibnamefont {Wu}}, \bibinfo {author} {\bibfnamefont {M.~J.}\ \bibnamefont {Collins}}, \bibinfo {author} {\bibfnamefont {C.}~\bibnamefont {Xiong}}, \bibinfo {author} {\bibfnamefont {A.}~\bibnamefont {Schreiber}}, \bibinfo {author} {\bibfnamefont {F.}~\bibnamefont {Katzschmann}}, \bibinfo {author} {\bibfnamefont {F.}~\bibnamefont {Eilenberger}}, \bibinfo {author} {\bibfnamefont {R.}~\bibnamefont {Schiek}},  \emph {et~al.},\ }\href@noop {} {\bibfield  {journal} {\bibinfo  {journal} {Physical Review X}\ }\textbf {\bibinfo {volume} {4}},\ \bibinfo {pages} {031007} (\bibinfo {year} {2014})}\BibitemShut {NoStop}%
\bibitem [{\citenamefont {Raymond}\ \emph {et~al.}(2024)\citenamefont {Raymond}, \citenamefont {Zecchetto}, \citenamefont {Palomo}, \citenamefont {Morassi}, \citenamefont {Lema{\^\i}tre}, \citenamefont {Raineri}, \citenamefont {Amanti}, \citenamefont {Ducci},\ and\ \citenamefont {Baboux}}]{raymond2024tunable}%
  \BibitemOpen
  \bibfield  {author} {\bibinfo {author} {\bibfnamefont {A.}~\bibnamefont {Raymond}}, \bibinfo {author} {\bibfnamefont {A.}~\bibnamefont {Zecchetto}}, \bibinfo {author} {\bibfnamefont {J.}~\bibnamefont {Palomo}}, \bibinfo {author} {\bibfnamefont {M.}~\bibnamefont {Morassi}}, \bibinfo {author} {\bibfnamefont {A.}~\bibnamefont {Lema{\^\i}tre}}, \bibinfo {author} {\bibfnamefont {F.}~\bibnamefont {Raineri}}, \bibinfo {author} {\bibfnamefont {M.}~\bibnamefont {Amanti}}, \bibinfo {author} {\bibfnamefont {S.}~\bibnamefont {Ducci}}, \ and\ \bibinfo {author} {\bibfnamefont {F.}~\bibnamefont {Baboux}},\ }\href@noop {} {\bibfield  {journal} {\bibinfo  {journal} {Physical Review Letters}\ }\textbf {\bibinfo {volume} {133}},\ \bibinfo {pages} {233602} (\bibinfo {year} {2024})}\BibitemShut {NoStop}%
\bibitem [{\citenamefont {Raymond}\ \emph {et~al.}(2025)\citenamefont {Raymond}, \citenamefont {Cathala}, \citenamefont {Morassi}, \citenamefont {Lemaitre}, \citenamefont {Raineri}, \citenamefont {Ducci},\ and\ \citenamefont {Baboux}}]{raymond2025tailoring}%
  \BibitemOpen
  \bibfield  {author} {\bibinfo {author} {\bibfnamefont {A.}~\bibnamefont {Raymond}}, \bibinfo {author} {\bibfnamefont {P.}~\bibnamefont {Cathala}}, \bibinfo {author} {\bibfnamefont {M.}~\bibnamefont {Morassi}}, \bibinfo {author} {\bibfnamefont {A.}~\bibnamefont {Lemaitre}}, \bibinfo {author} {\bibfnamefont {F.}~\bibnamefont {Raineri}}, \bibinfo {author} {\bibfnamefont {S.}~\bibnamefont {Ducci}}, \ and\ \bibinfo {author} {\bibfnamefont {F.}~\bibnamefont {Baboux}},\ }\href@noop {} {\bibfield  {journal} {\bibinfo  {journal} {Optics Express}\ }\textbf {\bibinfo {volume} {33}},\ \bibinfo {pages} {45869} (\bibinfo {year} {2025})}\BibitemShut {NoStop}%
\bibitem [{\citenamefont {Belsley}\ \emph {et~al.}(2020)\citenamefont {Belsley}, \citenamefont {Pertsch},\ and\ \citenamefont {Setzpfandt}}]{belsley2020generating}%
  \BibitemOpen
  \bibfield  {author} {\bibinfo {author} {\bibfnamefont {A.}~\bibnamefont {Belsley}}, \bibinfo {author} {\bibfnamefont {T.}~\bibnamefont {Pertsch}}, \ and\ \bibinfo {author} {\bibfnamefont {F.}~\bibnamefont {Setzpfandt}},\ }\href@noop {} {\bibfield  {journal} {\bibinfo  {journal} {Optics Express}\ }\textbf {\bibinfo {volume} {28}},\ \bibinfo {pages} {28792} (\bibinfo {year} {2020})}\BibitemShut {NoStop}%
\bibitem [{\citenamefont {Hamilton}\ \emph {et~al.}(2022)\citenamefont {Hamilton}, \citenamefont {Christ}, \citenamefont {Barkhofen}, \citenamefont {Barnett}, \citenamefont {Jex},\ and\ \citenamefont {Silberhorn}}]{hamilton2022quantum}%
  \BibitemOpen
  \bibfield  {author} {\bibinfo {author} {\bibfnamefont {C.~S.}\ \bibnamefont {Hamilton}}, \bibinfo {author} {\bibfnamefont {R.}~\bibnamefont {Christ}}, \bibinfo {author} {\bibfnamefont {S.}~\bibnamefont {Barkhofen}}, \bibinfo {author} {\bibfnamefont {S.~M.}\ \bibnamefont {Barnett}}, \bibinfo {author} {\bibfnamefont {I.}~\bibnamefont {Jex}}, \ and\ \bibinfo {author} {\bibfnamefont {C.}~\bibnamefont {Silberhorn}},\ }\href@noop {} {\bibfield  {journal} {\bibinfo  {journal} {Physical Review A}\ }\textbf {\bibinfo {volume} {105}},\ \bibinfo {pages} {042622} (\bibinfo {year} {2022})}\BibitemShut {NoStop}%
\bibitem [{\citenamefont {He}\ \emph {et~al.}(2024)\citenamefont {He}, \citenamefont {Xia}, \citenamefont {Leykam},\ and\ \citenamefont {Chen}}]{he2024optimizing}%
  \BibitemOpen
  \bibfield  {author} {\bibinfo {author} {\bibfnamefont {Y.}~\bibnamefont {He}}, \bibinfo {author} {\bibfnamefont {S.}~\bibnamefont {Xia}}, \bibinfo {author} {\bibfnamefont {D.}~\bibnamefont {Leykam}}, \ and\ \bibinfo {author} {\bibfnamefont {Z.}~\bibnamefont {Chen}},\ }\href@noop {} {\bibfield  {journal} {\bibinfo  {journal} {Optics Express}\ }\textbf {\bibinfo {volume} {32}},\ \bibinfo {pages} {32244} (\bibinfo {year} {2024})}\BibitemShut {NoStop}%
\bibitem [{\citenamefont {Delgado-Quesada}\ and\ \citenamefont {Rojas-Gonz{\~A}{\k{A}}lez}(2025)}]{delgado2025transport}%
  \BibitemOpen
  \bibfield  {author} {\bibinfo {author} {\bibfnamefont {J.}~\bibnamefont {Delgado-Quesada}}\ and\ \bibinfo {author} {\bibfnamefont {E.~A.}\ \bibnamefont {Rojas-Gonz{\~A}{\k{A}}lez}},\ }\href@noop {} {\bibfield  {journal} {\bibinfo  {journal} {arXiv preprint arXiv:2511.07273}\ } (\bibinfo {year} {2025})}\BibitemShut {NoStop}%
\bibitem [{\citenamefont {Longhi}(2007)}]{longhi2007light}%
  \BibitemOpen
  \bibfield  {author} {\bibinfo {author} {\bibfnamefont {S.}~\bibnamefont {Longhi}},\ }\href@noop {} {\bibfield  {journal} {\bibinfo  {journal} {Journal of Physics B: Atomic, Molecular and Optical Physics}\ }\textbf {\bibinfo {volume} {40}},\ \bibinfo {pages} {4477} (\bibinfo {year} {2007})}\BibitemShut {NoStop}%
\bibitem [{\citenamefont {Hudgings}\ \emph {et~al.}(2002)\citenamefont {Hudgings}, \citenamefont {Molter},\ and\ \citenamefont {Dutta}}]{hudgings2002design}%
  \BibitemOpen
  \bibfield  {author} {\bibinfo {author} {\bibfnamefont {J.}~\bibnamefont {Hudgings}}, \bibinfo {author} {\bibfnamefont {L.}~\bibnamefont {Molter}}, \ and\ \bibinfo {author} {\bibfnamefont {M.}~\bibnamefont {Dutta}},\ }\href@noop {} {\bibfield  {journal} {\bibinfo  {journal} {IEEE journal of quantum electronics}\ }\textbf {\bibinfo {volume} {36}},\ \bibinfo {pages} {1438} (\bibinfo {year} {2002})}\BibitemShut {NoStop}%
\bibitem [{\citenamefont {Hizanidis}\ \emph {et~al.}(2006)\citenamefont {Hizanidis}, \citenamefont {Droulias}, \citenamefont {Tsopelas}, \citenamefont {Efremidis},\ and\ \citenamefont {Christodoulides}}]{hizanidis2006localized}%
  \BibitemOpen
  \bibfield  {author} {\bibinfo {author} {\bibfnamefont {K.}~\bibnamefont {Hizanidis}}, \bibinfo {author} {\bibfnamefont {S.}~\bibnamefont {Droulias}}, \bibinfo {author} {\bibfnamefont {I.}~\bibnamefont {Tsopelas}}, \bibinfo {author} {\bibfnamefont {N.}~\bibnamefont {Efremidis}}, \ and\ \bibinfo {author} {\bibfnamefont {D.~N.}\ \bibnamefont {Christodoulides}},\ }\href@noop {} {\bibfield  {journal} {\bibinfo  {journal} {International Journal of Bifurcation and Chaos}\ }\textbf {\bibinfo {volume} {16}},\ \bibinfo {pages} {1739} (\bibinfo {year} {2006})}\BibitemShut {NoStop}%
\bibitem [{\citenamefont {Rai}\ and\ \citenamefont {Rai}(2022)}]{rai2022transfer}%
  \BibitemOpen
  \bibfield  {author} {\bibinfo {author} {\bibfnamefont {A.}~\bibnamefont {Rai}}\ and\ \bibinfo {author} {\bibfnamefont {A.}~\bibnamefont {Rai}},\ }\href@noop {} {\bibfield  {journal} {\bibinfo  {journal} {Journal of Optics}\ }\textbf {\bibinfo {volume} {24}},\ \bibinfo {pages} {125801} (\bibinfo {year} {2022})}\BibitemShut {NoStop}%
\bibitem [{\citenamefont {Lee}\ and\ \citenamefont {Park}(2018)}]{lee2018generation}%
  \BibitemOpen
  \bibfield  {author} {\bibinfo {author} {\bibfnamefont {H.~J.}\ \bibnamefont {Lee}}\ and\ \bibinfo {author} {\bibfnamefont {H.~S.}\ \bibnamefont {Park}},\ }\href@noop {} {\bibfield  {journal} {\bibinfo  {journal} {Photonics Research}\ }\textbf {\bibinfo {volume} {7}},\ \bibinfo {pages} {19} (\bibinfo {year} {2018})}\BibitemShut {NoStop}%
\bibitem [{\citenamefont {G{\'o}mez}\ \emph {et~al.}(2021)\citenamefont {G{\'o}mez}, \citenamefont {G{\'o}mez}, \citenamefont {Machuca}, \citenamefont {Cabello}, \citenamefont {P{\'a}dua}, \citenamefont {Walborn},\ and\ \citenamefont {Lima}}]{gomez2021multidimensional}%
  \BibitemOpen
  \bibfield  {author} {\bibinfo {author} {\bibfnamefont {E.~S.}\ \bibnamefont {G{\'o}mez}}, \bibinfo {author} {\bibfnamefont {S.}~\bibnamefont {G{\'o}mez}}, \bibinfo {author} {\bibfnamefont {I.}~\bibnamefont {Machuca}}, \bibinfo {author} {\bibfnamefont {A.}~\bibnamefont {Cabello}}, \bibinfo {author} {\bibfnamefont {S.}~\bibnamefont {P{\'a}dua}}, \bibinfo {author} {\bibfnamefont {S.}~\bibnamefont {Walborn}}, \ and\ \bibinfo {author} {\bibfnamefont {G.}~\bibnamefont {Lima}},\ }\href@noop {} {\bibfield  {journal} {\bibinfo  {journal} {Physical Review Applied}\ }\textbf {\bibinfo {volume} {15}},\ \bibinfo {pages} {034024} (\bibinfo {year} {2021})}\BibitemShut {NoStop}%
\bibitem [{\citenamefont {Barral}\ \emph {et~al.}(2019{\natexlab{a}})\citenamefont {Barral}, \citenamefont {Bencheikh}, \citenamefont {Belabas},\ and\ \citenamefont {Levenson}}]{barral2019zero}%
  \BibitemOpen
  \bibfield  {author} {\bibinfo {author} {\bibfnamefont {D.}~\bibnamefont {Barral}}, \bibinfo {author} {\bibfnamefont {K.}~\bibnamefont {Bencheikh}}, \bibinfo {author} {\bibfnamefont {N.}~\bibnamefont {Belabas}}, \ and\ \bibinfo {author} {\bibfnamefont {J.~A.}\ \bibnamefont {Levenson}},\ }\href@noop {} {\bibfield  {journal} {\bibinfo  {journal} {Physical Review A}\ }\textbf {\bibinfo {volume} {99}},\ \bibinfo {pages} {051801} (\bibinfo {year} {2019}{\natexlab{a}})}\BibitemShut {NoStop}%
\bibitem [{\citenamefont {Barral}\ \emph {et~al.}(2020{\natexlab{a}})\citenamefont {Barral}, \citenamefont {Walschaers}, \citenamefont {Bencheikh}, \citenamefont {Parigi}, \citenamefont {Levenson}, \citenamefont {Treps},\ and\ \citenamefont {Belabas}}]{barral2020versatile}%
  \BibitemOpen
  \bibfield  {author} {\bibinfo {author} {\bibfnamefont {D.}~\bibnamefont {Barral}}, \bibinfo {author} {\bibfnamefont {M.}~\bibnamefont {Walschaers}}, \bibinfo {author} {\bibfnamefont {K.}~\bibnamefont {Bencheikh}}, \bibinfo {author} {\bibfnamefont {V.}~\bibnamefont {Parigi}}, \bibinfo {author} {\bibfnamefont {J.~A.}\ \bibnamefont {Levenson}}, \bibinfo {author} {\bibfnamefont {N.}~\bibnamefont {Treps}}, \ and\ \bibinfo {author} {\bibfnamefont {N.}~\bibnamefont {Belabas}},\ }\href@noop {} {\bibfield  {journal} {\bibinfo  {journal} {Physical Review Applied}\ }\textbf {\bibinfo {volume} {14}},\ \bibinfo {pages} {044025} (\bibinfo {year} {2020}{\natexlab{a}})}\BibitemShut {NoStop}%
\bibitem [{\citenamefont {Rai}\ and\ \citenamefont {Angelakis}(2012)}]{rai2012dynamics}%
  \BibitemOpen
  \bibfield  {author} {\bibinfo {author} {\bibfnamefont {A.}~\bibnamefont {Rai}}\ and\ \bibinfo {author} {\bibfnamefont {D.~G.}\ \bibnamefont {Angelakis}},\ }\href@noop {} {\bibfield  {journal} {\bibinfo  {journal} {Physical Review A}\ }\textbf {\bibinfo {volume} {85}},\ \bibinfo {pages} {052330} (\bibinfo {year} {2012})}\BibitemShut {NoStop}%
\bibitem [{\citenamefont {Barral}\ \emph {et~al.}(2021{\natexlab{a}})\citenamefont {Barral}, \citenamefont {Bencheikh}, \citenamefont {Levenson},\ and\ \citenamefont {Belabas}}]{barral2021scalable}%
  \BibitemOpen
  \bibfield  {author} {\bibinfo {author} {\bibfnamefont {D.}~\bibnamefont {Barral}}, \bibinfo {author} {\bibfnamefont {K.}~\bibnamefont {Bencheikh}}, \bibinfo {author} {\bibfnamefont {J.~A.}\ \bibnamefont {Levenson}}, \ and\ \bibinfo {author} {\bibfnamefont {N.}~\bibnamefont {Belabas}},\ }\href@noop {} {\bibfield  {journal} {\bibinfo  {journal} {Physical Review Research}\ }\textbf {\bibinfo {volume} {3}},\ \bibinfo {pages} {013068} (\bibinfo {year} {2021}{\natexlab{a}})}\BibitemShut {NoStop}%
\bibitem [{\citenamefont {Anuradha}\ \emph {et~al.}(2024)\citenamefont {Anuradha}, \citenamefont {Patra}, \citenamefont {Gupta}, \citenamefont {Rai},\ and\ \citenamefont {Sen}}]{anuradha2024production}%
  \BibitemOpen
  \bibfield  {author} {\bibinfo {author} {\bibfnamefont {T.}~\bibnamefont {Anuradha}}, \bibinfo {author} {\bibfnamefont {A.}~\bibnamefont {Patra}}, \bibinfo {author} {\bibfnamefont {R.}~\bibnamefont {Gupta}}, \bibinfo {author} {\bibfnamefont {A.}~\bibnamefont {Rai}}, \ and\ \bibinfo {author} {\bibfnamefont {A.}~\bibnamefont {Sen}},\ }\href@noop {} {\bibfield  {journal} {\bibinfo  {journal} {Physical Review A}\ }\textbf {\bibinfo {volume} {109}},\ \bibinfo {pages} {032411} (\bibinfo {year} {2024})}\BibitemShut {NoStop}%
\bibitem [{\citenamefont {Meena}\ and\ \citenamefont {Rai}(2025)}]{meena2025theoretical}%
  \BibitemOpen
  \bibfield  {author} {\bibinfo {author} {\bibfnamefont {S.~S.}\ \bibnamefont {Meena}}\ and\ \bibinfo {author} {\bibfnamefont {A.}~\bibnamefont {Rai}},\ }\href@noop {} {\bibfield  {journal} {\bibinfo  {journal} {Journal of the Optical Society of America B}\ }\textbf {\bibinfo {volume} {42}},\ \bibinfo {pages} {825} (\bibinfo {year} {2025})}\BibitemShut {NoStop}%
\bibitem [{\citenamefont {Somekh}\ \emph {et~al.}(1973)\citenamefont {Somekh}, \citenamefont {Garmire}, \citenamefont {Yariv}, \citenamefont {Garvin},\ and\ \citenamefont {Hunsperger}}]{somekh1973channel}%
  \BibitemOpen
  \bibfield  {author} {\bibinfo {author} {\bibfnamefont {S.}~\bibnamefont {Somekh}}, \bibinfo {author} {\bibfnamefont {E.}~\bibnamefont {Garmire}}, \bibinfo {author} {\bibfnamefont {A.}~\bibnamefont {Yariv}}, \bibinfo {author} {\bibfnamefont {H.}~\bibnamefont {Garvin}}, \ and\ \bibinfo {author} {\bibfnamefont {R.}~\bibnamefont {Hunsperger}},\ }\href@noop {} {\bibfield  {journal} {\bibinfo  {journal} {Applied physics letters}\ }\textbf {\bibinfo {volume} {22}},\ \bibinfo {pages} {46} (\bibinfo {year} {1973})}\BibitemShut {NoStop}%
\bibitem [{\citenamefont {Alibart}\ \emph {et~al.}(2016)\citenamefont {Alibart}, \citenamefont {D’Auria}, \citenamefont {De~Micheli}, \citenamefont {Doutre}, \citenamefont {Kaiser}, \citenamefont {Labont{\'e}}, \citenamefont {Lunghi}, \citenamefont {Picholle},\ and\ \citenamefont {Tanzilli}}]{alibart2016quantum}%
  \BibitemOpen
  \bibfield  {author} {\bibinfo {author} {\bibfnamefont {O.}~\bibnamefont {Alibart}}, \bibinfo {author} {\bibfnamefont {V.}~\bibnamefont {D’Auria}}, \bibinfo {author} {\bibfnamefont {M.}~\bibnamefont {De~Micheli}}, \bibinfo {author} {\bibfnamefont {F.}~\bibnamefont {Doutre}}, \bibinfo {author} {\bibfnamefont {F.}~\bibnamefont {Kaiser}}, \bibinfo {author} {\bibfnamefont {L.}~\bibnamefont {Labont{\'e}}}, \bibinfo {author} {\bibfnamefont {T.}~\bibnamefont {Lunghi}}, \bibinfo {author} {\bibfnamefont {{\'E}.}~\bibnamefont {Picholle}}, \ and\ \bibinfo {author} {\bibfnamefont {S.}~\bibnamefont {Tanzilli}},\ }\href@noop {} {\bibfield  {journal} {\bibinfo  {journal} {Journal of Optics}\ }\textbf {\bibinfo {volume} {18}},\ \bibinfo {pages} {104001} (\bibinfo {year} {2016})}\BibitemShut {NoStop}%
\bibitem [{\citenamefont {Noda}\ \emph {et~al.}(1981)\citenamefont {Noda}, \citenamefont {Fukuma},\ and\ \citenamefont {Mikami}}]{noda1981design}%
  \BibitemOpen
  \bibfield  {author} {\bibinfo {author} {\bibfnamefont {J.}~\bibnamefont {Noda}}, \bibinfo {author} {\bibfnamefont {M.}~\bibnamefont {Fukuma}}, \ and\ \bibinfo {author} {\bibfnamefont {O.}~\bibnamefont {Mikami}},\ }\href@noop {} {\bibfield  {journal} {\bibinfo  {journal} {Applied Optics}\ }\textbf {\bibinfo {volume} {20}},\ \bibinfo {pages} {2284} (\bibinfo {year} {1981})}\BibitemShut {NoStop}%
\bibitem [{\citenamefont {Horoshko}(2022)}]{horoshko2022generator}%
  \BibitemOpen
  \bibfield  {author} {\bibinfo {author} {\bibfnamefont {D.~B.}\ \bibnamefont {Horoshko}},\ }\href@noop {} {\bibfield  {journal} {\bibinfo  {journal} {Physical Review A}\ }\textbf {\bibinfo {volume} {105}},\ \bibinfo {pages} {013708} (\bibinfo {year} {2022})}\BibitemShut {NoStop}%
\bibitem [{\citenamefont {Gray}(2006)}]{gray2006toeplitz}%
  \BibitemOpen
  \bibfield  {author} {\bibinfo {author} {\bibfnamefont {R.~M.}\ \bibnamefont {Gray}},\ }\href@noop {} {\bibfield  {journal} {\bibinfo  {journal} {Foundations and Trends in Communications and Information Theory}\ }\textbf {\bibinfo {volume} {2}},\ \bibinfo {pages} {155} (\bibinfo {year} {2006})}\BibitemShut {NoStop}%
\bibitem [{\citenamefont {Barral}\ \emph {et~al.}(2020{\natexlab{b}})\citenamefont {Barral}, \citenamefont {Walschaers}, \citenamefont {Bencheikh}, \citenamefont {Parigi}, \citenamefont {Levenson}, \citenamefont {Treps},\ and\ \citenamefont {Belabas}}]{barral2020quantum}%
  \BibitemOpen
  \bibfield  {author} {\bibinfo {author} {\bibfnamefont {D.}~\bibnamefont {Barral}}, \bibinfo {author} {\bibfnamefont {M.}~\bibnamefont {Walschaers}}, \bibinfo {author} {\bibfnamefont {K.}~\bibnamefont {Bencheikh}}, \bibinfo {author} {\bibfnamefont {V.}~\bibnamefont {Parigi}}, \bibinfo {author} {\bibfnamefont {J.~A.}\ \bibnamefont {Levenson}}, \bibinfo {author} {\bibfnamefont {N.}~\bibnamefont {Treps}}, \ and\ \bibinfo {author} {\bibfnamefont {N.}~\bibnamefont {Belabas}},\ }\href@noop {} {\bibfield  {journal} {\bibinfo  {journal} {Physical Review A}\ }\textbf {\bibinfo {volume} {102}},\ \bibinfo {pages} {043706} (\bibinfo {year} {2020}{\natexlab{b}})}\BibitemShut {NoStop}%
\bibitem [{\citenamefont {Fiur{\'a}{\v{s}}ek}\ and\ \citenamefont {Pe{\v{r}}ina}(2000)}]{fiuravsek2000substituting}%
  \BibitemOpen
  \bibfield  {author} {\bibinfo {author} {\bibfnamefont {J.}~\bibnamefont {Fiur{\'a}{\v{s}}ek}}\ and\ \bibinfo {author} {\bibfnamefont {J.}~\bibnamefont {Pe{\v{r}}ina}},\ }\href@noop {} {\bibfield  {journal} {\bibinfo  {journal} {Physical Review A}\ }\textbf {\bibinfo {volume} {62}},\ \bibinfo {pages} {033808} (\bibinfo {year} {2000})}\BibitemShut {NoStop}%
\bibitem [{\citenamefont {Mollow}\ and\ \citenamefont {Glauber}(1967)}]{mollow1967quantum}%
  \BibitemOpen
  \bibfield  {author} {\bibinfo {author} {\bibfnamefont {B.}~\bibnamefont {Mollow}}\ and\ \bibinfo {author} {\bibfnamefont {R.}~\bibnamefont {Glauber}},\ }\href@noop {} {\bibfield  {journal} {\bibinfo  {journal} {Physical Review}\ }\textbf {\bibinfo {volume} {160}},\ \bibinfo {pages} {1076} (\bibinfo {year} {1967})}\BibitemShut {NoStop}%
\bibitem [{\citenamefont {Adesso}\ \emph {et~al.}(2014)\citenamefont {Adesso}, \citenamefont {Ragy},\ and\ \citenamefont {Lee}}]{adesso2014continuous}%
  \BibitemOpen
  \bibfield  {author} {\bibinfo {author} {\bibfnamefont {G.}~\bibnamefont {Adesso}}, \bibinfo {author} {\bibfnamefont {S.}~\bibnamefont {Ragy}}, \ and\ \bibinfo {author} {\bibfnamefont {A.~R.}\ \bibnamefont {Lee}},\ }\href@noop {} {\bibfield  {journal} {\bibinfo  {journal} {Open Systems \& Information Dynamics}\ }\textbf {\bibinfo {volume} {21}},\ \bibinfo {pages} {1440001} (\bibinfo {year} {2014})}\BibitemShut {NoStop}%
\bibitem [{\citenamefont {Van~Loock}\ and\ \citenamefont {Furusawa}(2003)}]{van2003detecting}%
  \BibitemOpen
  \bibfield  {author} {\bibinfo {author} {\bibfnamefont {P.}~\bibnamefont {Van~Loock}}\ and\ \bibinfo {author} {\bibfnamefont {A.}~\bibnamefont {Furusawa}},\ }\href@noop {} {\bibfield  {journal} {\bibinfo  {journal} {Physical Review A}\ }\textbf {\bibinfo {volume} {67}},\ \bibinfo {pages} {052315} (\bibinfo {year} {2003})}\BibitemShut {NoStop}%
\bibitem [{\citenamefont {Braunstein}\ and\ \citenamefont {Van~Loock}(2005)}]{braunstein2005quantum}%
  \BibitemOpen
  \bibfield  {author} {\bibinfo {author} {\bibfnamefont {S.~L.}\ \bibnamefont {Braunstein}}\ and\ \bibinfo {author} {\bibfnamefont {P.}~\bibnamefont {Van~Loock}},\ }\href@noop {} {\bibfield  {journal} {\bibinfo  {journal} {Reviews of modern physics}\ }\textbf {\bibinfo {volume} {77}},\ \bibinfo {pages} {513} (\bibinfo {year} {2005})}\BibitemShut {NoStop}%
\bibitem [{\citenamefont {Barral}\ \emph {et~al.}(2021{\natexlab{b}})\citenamefont {Barral}, \citenamefont {D’auria}, \citenamefont {Doutre}, \citenamefont {Lunghi}, \citenamefont {Tanzilli}, \citenamefont {Petronela~Rambu}, \citenamefont {Tascu}, \citenamefont {Ariel~Levenson}, \citenamefont {Belabas},\ and\ \citenamefont {Bencheikh}}]{barral2021supermode}%
  \BibitemOpen
  \bibfield  {author} {\bibinfo {author} {\bibfnamefont {D.}~\bibnamefont {Barral}}, \bibinfo {author} {\bibfnamefont {V.}~\bibnamefont {D’auria}}, \bibinfo {author} {\bibfnamefont {F.}~\bibnamefont {Doutre}}, \bibinfo {author} {\bibfnamefont {T.}~\bibnamefont {Lunghi}}, \bibinfo {author} {\bibfnamefont {S.}~\bibnamefont {Tanzilli}}, \bibinfo {author} {\bibfnamefont {A.}~\bibnamefont {Petronela~Rambu}}, \bibinfo {author} {\bibfnamefont {S.}~\bibnamefont {Tascu}}, \bibinfo {author} {\bibfnamefont {J.}~\bibnamefont {Ariel~Levenson}}, \bibinfo {author} {\bibfnamefont {N.}~\bibnamefont {Belabas}}, \ and\ \bibinfo {author} {\bibfnamefont {K.}~\bibnamefont {Bencheikh}},\ }\href@noop {} {\bibfield  {journal} {\bibinfo  {journal} {Optics Express}\ }\textbf {\bibinfo {volume} {29}},\ \bibinfo {pages} {37175} (\bibinfo {year} {2021}{\natexlab{b}})}\BibitemShut {NoStop}%
\bibitem [{\citenamefont {Leonhardt}(1997)}]{leonhardt1997measuring}%
  \BibitemOpen
  \bibfield  {author} {\bibinfo {author} {\bibfnamefont {U.}~\bibnamefont {Leonhardt}},\ }\href@noop {} {\emph {\bibinfo {title} {Measuring the quantum state of light}}},\ Vol.~\bibinfo {volume} {22}\ (\bibinfo  {publisher} {Cambridge university press},\ \bibinfo {year} {1997})\BibitemShut {NoStop}%
\bibitem [{\citenamefont {Costas}\ \emph {et~al.}(2025)\citenamefont {Costas}, \citenamefont {Belabas},\ and\ \citenamefont {Barral}}]{costas2025spatio}%
  \BibitemOpen
  \bibfield  {author} {\bibinfo {author} {\bibfnamefont {N.}~\bibnamefont {Costas}}, \bibinfo {author} {\bibfnamefont {N.}~\bibnamefont {Belabas}}, \ and\ \bibinfo {author} {\bibfnamefont {D.}~\bibnamefont {Barral}},\ }\href@noop {} {\bibfield  {journal} {\bibinfo  {journal} {Physical Review Research}\ }\textbf {\bibinfo {volume} {7}},\ \bibinfo {pages} {043320} (\bibinfo {year} {2025})}\BibitemShut {NoStop}%
\bibitem [{\citenamefont {Heinrich}\ \emph {et~al.}(2008)\citenamefont {Heinrich}, \citenamefont {Szameit}, \citenamefont {Dreisow}, \citenamefont {D{\"o}ring}, \citenamefont {Thomas}, \citenamefont {Nolte}, \citenamefont {T{\"u}nnermann},\ and\ \citenamefont {Ancona}}]{heinrich2008evanescent}%
  \BibitemOpen
  \bibfield  {author} {\bibinfo {author} {\bibfnamefont {M.}~\bibnamefont {Heinrich}}, \bibinfo {author} {\bibfnamefont {A.}~\bibnamefont {Szameit}}, \bibinfo {author} {\bibfnamefont {F.}~\bibnamefont {Dreisow}}, \bibinfo {author} {\bibfnamefont {S.}~\bibnamefont {D{\"o}ring}}, \bibinfo {author} {\bibfnamefont {J.}~\bibnamefont {Thomas}}, \bibinfo {author} {\bibfnamefont {S.}~\bibnamefont {Nolte}}, \bibinfo {author} {\bibfnamefont {A.}~\bibnamefont {T{\"u}nnermann}}, \ and\ \bibinfo {author} {\bibfnamefont {A.}~\bibnamefont {Ancona}},\ }\href@noop {} {\bibfield  {journal} {\bibinfo  {journal} {Applied Physics Letters}\ }\textbf {\bibinfo {volume} {93}} (\bibinfo {year} {2008})}\BibitemShut {NoStop}%
\bibitem [{\citenamefont {Christodoulides}\ and\ \citenamefont {Eugenieva}(2001)}]{christodoulides2001blocking}%
  \BibitemOpen
  \bibfield  {author} {\bibinfo {author} {\bibfnamefont {D.~N.}\ \bibnamefont {Christodoulides}}\ and\ \bibinfo {author} {\bibfnamefont {E.~D.}\ \bibnamefont {Eugenieva}},\ }\href@noop {} {\bibfield  {journal} {\bibinfo  {journal} {Physical Review Letters}\ }\textbf {\bibinfo {volume} {87}},\ \bibinfo {pages} {233901} (\bibinfo {year} {2001})}\BibitemShut {NoStop}%
\bibitem [{\citenamefont {Szameit}\ and\ \citenamefont {Nolte}(2010)}]{szameit2010discrete}%
  \BibitemOpen
  \bibfield  {author} {\bibinfo {author} {\bibfnamefont {A.}~\bibnamefont {Szameit}}\ and\ \bibinfo {author} {\bibfnamefont {S.}~\bibnamefont {Nolte}},\ }\href@noop {} {\bibfield  {journal} {\bibinfo  {journal} {Journal of Physics B: Atomic, Molecular and Optical Physics}\ }\textbf {\bibinfo {volume} {43}},\ \bibinfo {pages} {163001} (\bibinfo {year} {2010})}\BibitemShut {NoStop}%
\bibitem [{\citenamefont {Meany}\ \emph {et~al.}(2015)\citenamefont {Meany}, \citenamefont {Gr{\"a}fe}, \citenamefont {Heilmann}, \citenamefont {Perez-Leija}, \citenamefont {Gross}, \citenamefont {Steel}, \citenamefont {Withford},\ and\ \citenamefont {Szameit}}]{meany2015laser}%
  \BibitemOpen
  \bibfield  {author} {\bibinfo {author} {\bibfnamefont {T.}~\bibnamefont {Meany}}, \bibinfo {author} {\bibfnamefont {M.}~\bibnamefont {Gr{\"a}fe}}, \bibinfo {author} {\bibfnamefont {R.}~\bibnamefont {Heilmann}}, \bibinfo {author} {\bibfnamefont {A.}~\bibnamefont {Perez-Leija}}, \bibinfo {author} {\bibfnamefont {S.}~\bibnamefont {Gross}}, \bibinfo {author} {\bibfnamefont {M.~J.}\ \bibnamefont {Steel}}, \bibinfo {author} {\bibfnamefont {M.~J.}\ \bibnamefont {Withford}}, \ and\ \bibinfo {author} {\bibfnamefont {A.}~\bibnamefont {Szameit}},\ }\href@noop {} {\bibfield  {journal} {\bibinfo  {journal} {Laser \& Photonics Reviews}\ }\textbf {\bibinfo {volume} {9}},\ \bibinfo {pages} {363} (\bibinfo {year} {2015})}\BibitemShut {NoStop}%
\bibitem [{\citenamefont {Li}\ \emph {et~al.}(2022{\natexlab{b}})\citenamefont {Li}, \citenamefont {Kong},\ and\ \citenamefont {Chen}}]{li2022femtosecond}%
  \BibitemOpen
  \bibfield  {author} {\bibinfo {author} {\bibfnamefont {L.}~\bibnamefont {Li}}, \bibinfo {author} {\bibfnamefont {W.}~\bibnamefont {Kong}}, \ and\ \bibinfo {author} {\bibfnamefont {F.}~\bibnamefont {Chen}},\ }\href@noop {} {\bibfield  {journal} {\bibinfo  {journal} {Advanced Photonics}\ }\textbf {\bibinfo {volume} {4}},\ \bibinfo {pages} {024002} (\bibinfo {year} {2022}{\natexlab{b}})}\BibitemShut {NoStop}%
\bibitem [{\citenamefont {Xu}\ \emph {et~al.}(2022)\citenamefont {Xu}, \citenamefont {Wang}, \citenamefont {Chen}, \citenamefont {Zhou}, \citenamefont {Ma}, \citenamefont {Wei}, \citenamefont {Wang}, \citenamefont {Niu}, \citenamefont {Fang}, \citenamefont {Wu} \emph {et~al.}}]{xu2022femtosecond}%
  \BibitemOpen
  \bibfield  {author} {\bibinfo {author} {\bibfnamefont {X.}~\bibnamefont {Xu}}, \bibinfo {author} {\bibfnamefont {T.}~\bibnamefont {Wang}}, \bibinfo {author} {\bibfnamefont {P.}~\bibnamefont {Chen}}, \bibinfo {author} {\bibfnamefont {C.}~\bibnamefont {Zhou}}, \bibinfo {author} {\bibfnamefont {J.}~\bibnamefont {Ma}}, \bibinfo {author} {\bibfnamefont {D.}~\bibnamefont {Wei}}, \bibinfo {author} {\bibfnamefont {H.}~\bibnamefont {Wang}}, \bibinfo {author} {\bibfnamefont {B.}~\bibnamefont {Niu}}, \bibinfo {author} {\bibfnamefont {X.}~\bibnamefont {Fang}}, \bibinfo {author} {\bibfnamefont {D.}~\bibnamefont {Wu}},  \emph {et~al.},\ }\href@noop {} {\bibfield  {journal} {\bibinfo  {journal} {Nature}\ }\textbf {\bibinfo {volume} {609}},\ \bibinfo {pages} {496} (\bibinfo {year} {2022})}\BibitemShut {NoStop}%
\bibitem [{\citenamefont {Jin}\ \emph {et~al.}(2014)\citenamefont {Jin}, \citenamefont {Liu}, \citenamefont {Xu}, \citenamefont {Xia}, \citenamefont {Zhong}, \citenamefont {Yuan}, \citenamefont {Zhou}, \citenamefont {Gong}, \citenamefont {Wang},\ and\ \citenamefont {Zhu}}]{jin2014chip}%
  \BibitemOpen
  \bibfield  {author} {\bibinfo {author} {\bibfnamefont {H.}~\bibnamefont {Jin}}, \bibinfo {author} {\bibfnamefont {F.}~\bibnamefont {Liu}}, \bibinfo {author} {\bibfnamefont {P.}~\bibnamefont {Xu}}, \bibinfo {author} {\bibfnamefont {J.}~\bibnamefont {Xia}}, \bibinfo {author} {\bibfnamefont {M.}~\bibnamefont {Zhong}}, \bibinfo {author} {\bibfnamefont {Y.}~\bibnamefont {Yuan}}, \bibinfo {author} {\bibfnamefont {J.}~\bibnamefont {Zhou}}, \bibinfo {author} {\bibfnamefont {Y.}~\bibnamefont {Gong}}, \bibinfo {author} {\bibfnamefont {W.}~\bibnamefont {Wang}}, \ and\ \bibinfo {author} {\bibfnamefont {S.}~\bibnamefont {Zhu}},\ }\href@noop {} {\bibfield  {journal} {\bibinfo  {journal} {Physical review letters}\ }\textbf {\bibinfo {volume} {113}},\ \bibinfo {pages} {103601} (\bibinfo {year} {2014})}\BibitemShut {NoStop}%
\bibitem [{\citenamefont {Mondain}\ \emph {et~al.}(2019)\citenamefont {Mondain}, \citenamefont {Lunghi}, \citenamefont {Zavatta}, \citenamefont {Gouzien}, \citenamefont {Doutre}, \citenamefont {De~Micheli}, \citenamefont {Tanzilli},\ and\ \citenamefont {D’Auria}}]{mondain2019chip}%
  \BibitemOpen
  \bibfield  {author} {\bibinfo {author} {\bibfnamefont {F.}~\bibnamefont {Mondain}}, \bibinfo {author} {\bibfnamefont {T.}~\bibnamefont {Lunghi}}, \bibinfo {author} {\bibfnamefont {A.}~\bibnamefont {Zavatta}}, \bibinfo {author} {\bibfnamefont {E.}~\bibnamefont {Gouzien}}, \bibinfo {author} {\bibfnamefont {F.}~\bibnamefont {Doutre}}, \bibinfo {author} {\bibfnamefont {M.}~\bibnamefont {De~Micheli}}, \bibinfo {author} {\bibfnamefont {S.}~\bibnamefont {Tanzilli}}, \ and\ \bibinfo {author} {\bibfnamefont {V.}~\bibnamefont {D’Auria}},\ }\href@noop {} {\bibfield  {journal} {\bibinfo  {journal} {Photonics Research}\ }\textbf {\bibinfo {volume} {7}},\ \bibinfo {pages} {A36} (\bibinfo {year} {2019})}\BibitemShut {NoStop}%
\bibitem [{\citenamefont {Osellame}\ \emph {et~al.}(2007)\citenamefont {Osellame}, \citenamefont {Lobino}, \citenamefont {Chiodo}, \citenamefont {Marangoni}, \citenamefont {Cerullo}, \citenamefont {Ramponi}, \citenamefont {Bookey}, \citenamefont {Thomson}, \citenamefont {Psaila},\ and\ \citenamefont {Kar}}]{osellame2007femtosecond}%
  \BibitemOpen
  \bibfield  {author} {\bibinfo {author} {\bibfnamefont {R.}~\bibnamefont {Osellame}}, \bibinfo {author} {\bibfnamefont {M.}~\bibnamefont {Lobino}}, \bibinfo {author} {\bibfnamefont {N.}~\bibnamefont {Chiodo}}, \bibinfo {author} {\bibfnamefont {M.}~\bibnamefont {Marangoni}}, \bibinfo {author} {\bibfnamefont {G.}~\bibnamefont {Cerullo}}, \bibinfo {author} {\bibfnamefont {R.}~\bibnamefont {Ramponi}}, \bibinfo {author} {\bibfnamefont {H.~T.}\ \bibnamefont {Bookey}}, \bibinfo {author} {\bibfnamefont {R.~R.}\ \bibnamefont {Thomson}}, \bibinfo {author} {\bibfnamefont {N.~D.}\ \bibnamefont {Psaila}}, \ and\ \bibinfo {author} {\bibfnamefont {A.~K.}\ \bibnamefont {Kar}},\ }\href@noop {} {\bibfield  {journal} {\bibinfo  {journal} {Applied physics letters}\ }\textbf {\bibinfo {volume} {90}} (\bibinfo {year} {2007})}\BibitemShut {NoStop}%
\bibitem [{\citenamefont {L{\"u}}\ \emph {et~al.}(2023)\citenamefont {L{\"u}}, \citenamefont {Li}, \citenamefont {Ma},\ and\ \citenamefont {Chen}}]{lu2023femtosecond}%
  \BibitemOpen
  \bibfield  {author} {\bibinfo {author} {\bibfnamefont {J.}~\bibnamefont {L{\"u}}}, \bibinfo {author} {\bibfnamefont {G.}~\bibnamefont {Li}}, \bibinfo {author} {\bibfnamefont {Y.}~\bibnamefont {Ma}}, \ and\ \bibinfo {author} {\bibfnamefont {F.}~\bibnamefont {Chen}},\ }\href@noop {} {\bibfield  {journal} {\bibinfo  {journal} {Chinese Optics Letters}\ }\textbf {\bibinfo {volume} {21}},\ \bibinfo {pages} {112201} (\bibinfo {year} {2023})}\BibitemShut {NoStop}%
\bibitem [{\citenamefont {Barral}\ \emph {et~al.}(2019{\natexlab{b}})\citenamefont {Barral}, \citenamefont {Belabas}, \citenamefont {Bencheikh},\ and\ \citenamefont {Levenson}}]{barral2019coupling}%
  \BibitemOpen
  \bibfield  {author} {\bibinfo {author} {\bibfnamefont {D.}~\bibnamefont {Barral}}, \bibinfo {author} {\bibfnamefont {N.}~\bibnamefont {Belabas}}, \bibinfo {author} {\bibfnamefont {K.}~\bibnamefont {Bencheikh}}, \ and\ \bibinfo {author} {\bibfnamefont {J.~A.}\ \bibnamefont {Levenson}},\ }\href@noop {} {\bibfield  {journal} {\bibinfo  {journal} {Physical Review A}\ }\textbf {\bibinfo {volume} {100}},\ \bibinfo {pages} {013824} (\bibinfo {year} {2019}{\natexlab{b}})}\BibitemShut {NoStop}%
\bibitem [{\citenamefont {Due{\~n}as}\ \emph {et~al.}(2021)\citenamefont {Due{\~n}as}, \citenamefont {P{\'e}rez}, \citenamefont {Hermann-Avigliano},\ and\ \citenamefont {Torres}}]{duenas2021quadrature}%
  \BibitemOpen
  \bibfield  {author} {\bibinfo {author} {\bibfnamefont {J.~M.}\ \bibnamefont {Due{\~n}as}}, \bibinfo {author} {\bibfnamefont {G.~O.}\ \bibnamefont {P{\'e}rez}}, \bibinfo {author} {\bibfnamefont {C.}~\bibnamefont {Hermann-Avigliano}}, \ and\ \bibinfo {author} {\bibfnamefont {L.~E.~F.}\ \bibnamefont {Torres}},\ }\href@noop {} {\bibfield  {journal} {\bibinfo  {journal} {Quantum}\ }\textbf {\bibinfo {volume} {5}},\ \bibinfo {pages} {526} (\bibinfo {year} {2021})}\BibitemShut {NoStop}%
\bibitem [{\citenamefont {Kokkinakis}\ \emph {et~al.}(2024)\citenamefont {Kokkinakis}, \citenamefont {Makris},\ and\ \citenamefont {Economou}}]{kokkinakis2024anderson}%
  \BibitemOpen
  \bibfield  {author} {\bibinfo {author} {\bibfnamefont {E.}~\bibnamefont {Kokkinakis}}, \bibinfo {author} {\bibfnamefont {K.}~\bibnamefont {Makris}}, \ and\ \bibinfo {author} {\bibfnamefont {E.}~\bibnamefont {Economou}},\ }\href@noop {} {\bibfield  {journal} {\bibinfo  {journal} {Physical Review A}\ }\textbf {\bibinfo {volume} {110}},\ \bibinfo {pages} {053517} (\bibinfo {year} {2024})}\BibitemShut {NoStop}%
\bibitem [{\citenamefont {Ren}\ \emph {et~al.}(2022)\citenamefont {Ren}, \citenamefont {Lu}, \citenamefont {Jiang}, \citenamefont {Gao}, \citenamefont {Zhou}, \citenamefont {Wang}, \citenamefont {Jiao}, \citenamefont {Wang}, \citenamefont {Solntsev},\ and\ \citenamefont {Jin}}]{ren2022topologically}%
  \BibitemOpen
  \bibfield  {author} {\bibinfo {author} {\bibfnamefont {R.-J.}\ \bibnamefont {Ren}}, \bibinfo {author} {\bibfnamefont {Y.-H.}\ \bibnamefont {Lu}}, \bibinfo {author} {\bibfnamefont {Z.-K.}\ \bibnamefont {Jiang}}, \bibinfo {author} {\bibfnamefont {J.}~\bibnamefont {Gao}}, \bibinfo {author} {\bibfnamefont {W.-H.}\ \bibnamefont {Zhou}}, \bibinfo {author} {\bibfnamefont {Y.}~\bibnamefont {Wang}}, \bibinfo {author} {\bibfnamefont {Z.-Q.}\ \bibnamefont {Jiao}}, \bibinfo {author} {\bibfnamefont {X.-W.}\ \bibnamefont {Wang}}, \bibinfo {author} {\bibfnamefont {A.~S.}\ \bibnamefont {Solntsev}}, \ and\ \bibinfo {author} {\bibfnamefont {X.-M.}\ \bibnamefont {Jin}},\ }\href@noop {} {\bibfield  {journal} {\bibinfo  {journal} {Photonics Research}\ }\textbf {\bibinfo {volume} {10}},\ \bibinfo {pages} {456} (\bibinfo {year} {2022})}\BibitemShut {NoStop}%
\bibitem [{\citenamefont {Zecchetto}\ \emph {et~al.}(2025)\citenamefont {Zecchetto}, \citenamefont {Coudevylle}, \citenamefont {Morassi}, \citenamefont {Lema{\~A}{\v{Z}}tre}, \citenamefont {Amanti}, \citenamefont {Ducci},\ and\ \citenamefont {Baboux}}]{zecchetto2025topological}%
  \BibitemOpen
  \bibfield  {author} {\bibinfo {author} {\bibfnamefont {A.}~\bibnamefont {Zecchetto}}, \bibinfo {author} {\bibfnamefont {J.-R.}\ \bibnamefont {Coudevylle}}, \bibinfo {author} {\bibfnamefont {M.}~\bibnamefont {Morassi}}, \bibinfo {author} {\bibfnamefont {A.}~\bibnamefont {Lema{\~A}{\v{Z}}tre}}, \bibinfo {author} {\bibfnamefont {M.}~\bibnamefont {Amanti}}, \bibinfo {author} {\bibfnamefont {S.}~\bibnamefont {Ducci}}, \ and\ \bibinfo {author} {\bibfnamefont {F.}~\bibnamefont {Baboux}},\ }\href@noop {} {\bibfield  {journal} {\bibinfo  {journal} {arXiv preprint arXiv:2510.23796}\ } (\bibinfo {year} {2025})}\BibitemShut {NoStop}%
\bibitem [{\citenamefont {Hu}\ \emph {et~al.}(2023)\citenamefont {Hu}, \citenamefont {Guo}, \citenamefont {Liu}, \citenamefont {Li},\ and\ \citenamefont {Guo}}]{hu2023progress}%
  \BibitemOpen
  \bibfield  {author} {\bibinfo {author} {\bibfnamefont {X.-M.}\ \bibnamefont {Hu}}, \bibinfo {author} {\bibfnamefont {Y.}~\bibnamefont {Guo}}, \bibinfo {author} {\bibfnamefont {B.-H.}\ \bibnamefont {Liu}}, \bibinfo {author} {\bibfnamefont {C.-F.}\ \bibnamefont {Li}}, \ and\ \bibinfo {author} {\bibfnamefont {G.-C.}\ \bibnamefont {Guo}},\ }\href@noop {} {\bibfield  {journal} {\bibinfo  {journal} {Nature Reviews Physics}\ }\textbf {\bibinfo {volume} {5}},\ \bibinfo {pages} {339} (\bibinfo {year} {2023})}\BibitemShut {NoStop}%
\bibitem [{\citenamefont {Walmsley}(2023)}]{walmsley2023light}%
  \BibitemOpen
  \bibfield  {author} {\bibinfo {author} {\bibfnamefont {I.}~\bibnamefont {Walmsley}},\ }\href@noop {} {\bibfield  {journal} {\bibinfo  {journal} {Optica Quantum}\ }\textbf {\bibinfo {volume} {1}},\ \bibinfo {pages} {35} (\bibinfo {year} {2023})}\BibitemShut {NoStop}%
\bibitem [{\citenamefont {Luo}\ \emph {et~al.}(2023)\citenamefont {Luo}, \citenamefont {Cao}, \citenamefont {Shi}, \citenamefont {Wan}, \citenamefont {Zhang}, \citenamefont {Li}, \citenamefont {Chen}, \citenamefont {Li}, \citenamefont {Li}, \citenamefont {Wang} \emph {et~al.}}]{luo2023recent}%
  \BibitemOpen
  \bibfield  {author} {\bibinfo {author} {\bibfnamefont {W.}~\bibnamefont {Luo}}, \bibinfo {author} {\bibfnamefont {L.}~\bibnamefont {Cao}}, \bibinfo {author} {\bibfnamefont {Y.}~\bibnamefont {Shi}}, \bibinfo {author} {\bibfnamefont {L.}~\bibnamefont {Wan}}, \bibinfo {author} {\bibfnamefont {H.}~\bibnamefont {Zhang}}, \bibinfo {author} {\bibfnamefont {S.}~\bibnamefont {Li}}, \bibinfo {author} {\bibfnamefont {G.}~\bibnamefont {Chen}}, \bibinfo {author} {\bibfnamefont {Y.}~\bibnamefont {Li}}, \bibinfo {author} {\bibfnamefont {S.}~\bibnamefont {Li}}, \bibinfo {author} {\bibfnamefont {Y.}~\bibnamefont {Wang}},  \emph {et~al.},\ }\href@noop {} {\bibfield  {journal} {\bibinfo  {journal} {Light: Science \& Applications}\ }\textbf {\bibinfo {volume} {12}},\ \bibinfo {pages} {175} (\bibinfo {year} {2023})}\BibitemShut {NoStop}%
\bibitem [{\citenamefont {Barbieri}(2022)}]{barbieri2022optical}%
  \BibitemOpen
  \bibfield  {author} {\bibinfo {author} {\bibfnamefont {M.}~\bibnamefont {Barbieri}},\ }\href@noop {} {\bibfield  {journal} {\bibinfo  {journal} {PRX Quantum}\ }\textbf {\bibinfo {volume} {3}},\ \bibinfo {pages} {010202} (\bibinfo {year} {2022})}\BibitemShut {NoStop}%
\end{thebibliography}%

\end{document}